\documentclass[
aps,
reprint,
nofootinbib,
longbibliography,
superscriptaddress
]{revtex4-2}

\usepackage[english]{babel}
\usepackage[utf8x]{inputenc}
\usepackage[T1]{fontenc}

\usepackage{tikz}
\usepackage{float}
\usepackage{amsmath}
\usepackage{amssymb}
\usepackage{booktabs}
\usepackage{derivative}

\usepackage{hyperref}
\usepackage[capitalize]{cleveref}
\usepackage{graphicx}
\usepackage[indexonlyfirst, nogroupskip]{glossaries}
\usepackage{physics}
\usepackage{siunitx}
\usepackage{tikz-3dplot}
\usepackage[normalem]{ulem} % <----- This can be removed. Used for crossing out phrases 

\makeglossaries
\newacronym{lisa}{LISA}{Laser Interferometer Space Antenna}
\newacronym{esa}{ESA}{European Space Agency}

\newacronym{gw}{GW}{gravitational wave}
\newacronym{sgw}{SGW}{stochastic gravitational wave}
\newacronym{sgwb}{SGWB}{stochastic gravitational-wave background}
\newacronym{gr}{GR}{general relativity}

\newacronym{inrep}{INReP}{initial noise reduction pipeline}
\newacronym{tdi}{TDI}{time-delay interferometry}
\newacronym{mosa}{MOSA}{movable optical sub-assembly}
\newacronym{oms}{OMS}{optical metrology system}

\newacronym[longplural={power spectral densities}]{psd}{PSD}{power spectral density}
\newacronym[longplural={amplitude spectral densities}]{asd}{ASD}{amplitude spectral density}
\newacronym[longplural={cross spectral densities}]{csd}{CSD}{cross spectral density}
\newacronym{rms}{RMS}{root mean square}
\newacronym{snr}{SNR}{signal-to-noise ratio}
\newacronym[longplural={discrete Fourier transforms}]{dft}{DFT}{discrete Fourier transform}

\newacronym{map}{MAP}{maximum a posteriori estimate}
\newacronym{dof}{DoF}{degrees of freedom}
\newacronym{mcmc}{MCMC}{Markov chain Monte Carlo}
\newacronym{rj}{RJ}{reversible jump}

\newacronym{pta}{PTA}{pulsar timing array}
\newacronym{ipta}{IPTA}{International Pulsar Timing Array}
\tdplotsetmaincoords{60}{110}
\pgfmathsetmacro{\rvec}{1.0}
\pgfmathsetmacro{\thetavec}{30}
\pgfmathsetmacro{\phivec}{60}

\newcommand{\delay}[1]{\mathbf{D}_{#1}}

\definecolor{blue}{rgb}{0.3, 0.4, 0.8}
\definecolor{amaranth}{rgb}{0.9, 0.17, 0.31}
\definecolor{pink}{rgb}{0.87, 0.56, 0.81}
\definecolor{ao}{rgb}{0.0, 0.5, 0.0}
\definecolor{maroon}{rgb}{0.76, 0.13, 0.28}
\definecolor{cardinal}{rgb}{0.77, 0.12, 0.23}
\definecolor{yellow}{rgb}{1.0, 1.0, 0.87}
\definecolor{lightseagreen}{rgb}{0.45, 0.85, 0.58}
\definecolor{gray}{rgb}{0.9, 0.9, 0.9}
\definecolor{lightblue}{rgb}{0.66, 0.84, 0.96}

\hypersetup{
    citecolor=blue,
    colorlinks=true,
    linkcolor=blue,
    urlcolor=blue
}

\DeclareSIUnit\year{yr}

\begin{document}
	
	\preprint{APS/123-QED}
	\title{Uncovering gravitational-wave backgrounds from noises of unknown shape with LISA}
	
	\author{Quentin Baghi}
	\email{quentin.baghi@cea.fr}
	\affiliation{CEA Paris-Saclay University, Irfu/DPhP, Bat. 141, 91191 Gif sur Yvette Cedex, France}
	
	\author{Nikolaos Karnesis}
	\affiliation{Department of Physics, Aristotle University of Thessaloniki, Thessaloniki 54124, Greece}
	
	\author{Jean-Baptiste Bayle}
	\affiliation{University of Glasgow, Glasgow G12 8QQ, United Kingdom}
	
	\author{Marc Besan\c{c}on}
	\affiliation{CEA Paris-Saclay University, Irfu/DPhP, Bat. 141, 91191 Gif sur Yvette Cedex, France}
	
	\author{Henri Inchauspé}
	\affiliation{Institut für Theoretische Physik, Universität Heidelberg, Philosophenweg 16, 69120 Heidelberg, Germany}
	
	\date{\today}
	

\begin{abstract}
Detecting stochastic background radiation of cosmological origin is an exciting possibility for current and future gravitational-wave (GW) detectors. However, distinguishing it from other stochastic processes, such as instrumental noise and astrophysical backgrounds, is challenging. It is even more delicate for the space-based GW observatory LISA since it cannot correlate its observations with other detectors, unlike today’s terrestrial network. Nonetheless, with multiple measurements across the constellation and high accuracy in the noise level, detection is still possible. In the context of GW background detection, previous studies have assumed that instrumental noise has a known, possibly parameterized, spectral shape. To make our analysis robust against imperfect knowledge of the instrumental noise, we challenge this crucial assumption and assume that the single-link interferometric noises have an arbitrary and unknown spectrum. We investigate possible ways of separating instrumental and GW contributions by using realistic LISA data simulations with time-varying arms and second-generation time-delay interferometry. By fitting a generic spline model to the interferometer noise and a power-law template to the signal, we can detect GW stochastic backgrounds up to energy density levels comparable with fixed-shape models. We also demonstrate that we can probe a region of the GW background parameter space that today’s detectors cannot access.
\end{abstract}

%%% 3D-plot settings
%\tdplotsetmaincoords{60}{110}
%\pgfmathsetmacro{\rvec}{1.0}
%\pgfmathsetmacro{\thetavec}{30}
%\pgfmathsetmacro{\phivec}{60}
%
%
%% Define some nice colors
%\definecolor{blue}{rgb}{0.3, 0.4, 0.8}
%\definecolor{amaranth}{rgb}{0.9, 0.17, 0.31}
%\definecolor{pink}{rgb}{0.87, 0.56, 0.81}
%\definecolor{ao}{rgb}{0.0, 0.5, 0.0}
%\definecolor{maroon}{rgb}{0.76, 0.13, 0.28}
%\definecolor{cardinal}{rgb}{0.77, 0.12, 0.23}
%\definecolor{yellow}{rgb}{1.0, 1.0, 0.87}
%\definecolor{lightseagreen}{rgb}{0.45, 0.85, 0.58}
%\definecolor{gray}{rgb}{0.9, 0.9, 0.9}
%\definecolor{lightblue}{rgb}{0.66, 0.84, 0.96}
%
%% Change color of glossary links
%\renewcommand*{\glstextformat}[1]{\textcolor{blue}{#1}}
%
%% Change color of hyperref links
%\hypersetup{
%	citecolor=blue,
%	colorlinks=true,
%	linkcolor=blue,
%	urlcolor=blue
%}
%% SI units
%\DeclareSIUnit\year{yr}
%% ===================================================================

\glsresetall
\maketitle

%\flushbottom
%\glsresetall

\section{Introduction}
\label{sec:intro}

The hunt for \glspl{sgwb} (see~\cite{Regimbau2011rp, romano_detection_2017, Maggiore2018sht, Caprini2018mtu, Christensen_2019, Renzini2022, remortel_stochastic_2023} for recent reviews) has started with the advent of \gls{gw} astronomy, based on sensitive laser interferometry~\cite{LIGOScientific2016nwa,LIGOScientific2016jlg,LIGOScientific2017zlf,LIGOScientific2017ikf,LIGOScientific2018czr,LIGOScientific2019vic,LIGOScientific2019gaw,LIGOScientific2021nrg,KAGRA:2021kbb,KAGRA:2021mth,LIGOScientific2021oez} and the pulsar timing arrays~\cite{NANOGrav2015aud,NANOGRAV2018hou,NANOGrav2019ydy,NANOGrav2020bcs,NANOGrav2021flc,Antoniadis:2022pcn}. Future earth-based experiments~\cite{LIGOScientific2014pky,LIGOScientific2016wof,Maggiore2019uih} as well as space-borne missions~\cite{Bender2013nsa,Baker2019pnp,Sedda2019uro,Kawamura2011zz,Sato2017dkf,decigo,Baibhav2019rsa,Sesana2019vho} will also join this hunt. For the \gls{lisa} mission~\cite{LISA2019} in particular, the search for a \gls{sgwb} constitutes a major science objective. 

Produced by multiple incoherent emissions, stochastic \glspl{gw} can stem from both cosmological and astrophysical origins. In cosmology, they could originate for primordial quantum fluctuations possibly amplified by the cosmic inflation. They would then be unique tracers of the early and opaque universe, well before the last scattering surface. Other mechanisms like first-order phase transitions and cosmic strings, could also produce stochastic emissions of \glspl{gw}, carrying information about the existence of topological defects in the early universe. Thus the detection of \gls{sgwb} by \gls{lisa} should provide invaluable information on the astrophysical sources properties and could give hints on some of the physics processes which may have taken place in the early universe. However, in order to carry out this scientific program, it will be mandatory to be able to distinguish the sources signal from the instrumental background noise, which represents a major challenge for \gls{lisa}. Sorting out the sources categories in order to shed light on the underlying physics of a cosmological \gls{sgwb} represents yet an additional major challenge. 

The precise shape of the \gls{sgwb} spectrum from cosmological origin over the entire \gls{lisa} frequency band is difficult to predict can be considered unknown at present time. A wide variety of possible early-universe phenomena, either at the inflationary or post-inflationary stages, are possible source candidates. Likewise, large numbers of uncorrelated and unresolved astrophysical sources can superimpose and lead to \glspl{sgwb} with complex spectral shapes. \Glspl{sgwb} from both cosmological and astrophysical origins are furthermore likely to overlap, thus resulting in a \gls{sgwb} even more complex to decipher thus yielding a complex total \gls{sgwb} which would be challenging to characterize.

Capturing the main features of a \gls{sgwb} spectral shape and identifying its origin using parametrizations with various level of complexity is therefore a challenging task. Widely used parametrizations include simple power laws, monotonic signals with smoothly growing or decreasing slopes, signals with one or more exponential bumps, broken power laws, given by smooth function with changing slope at some given frequencies, or wiggly signals. Among the many challenges of dealing with \gls{sgwb}, assessing \gls{lisa}'s capability to separate different components, i.e., instrumental noise, galactic and extra-galactic foregrounds, astrophysical backgrounds, as well cosmological backgrounds, is of particular importance. Much work in these two directions has already begun (see for example~\cite{Cornish2001bb,Adams2010vc,Adams2013qma,Cornish2015pda,Parida2019ybm,Karnesis2019mph,Caprini2019pxz,Smith2019wny,Pieroni2020rob,Flauger2020qyi,Boileau2020rpg,Karnesis2021tsh,Boileau2022strings}). 

In contrast to previous search methods where the \gls{lisa} instrumental noise was parametrized with a fixed and known spectral shape, we investigate in this paper an approach to distinguish a simple \gls{sgwb} signal from the instrumental noise assuming that the single-link interferometric noises have an arbitrary and unknown spectrum. As a proof of principle, we choose to restrict ourselves to simple power laws to describe the \gls{sgwb} signal, deferring the discussion of more complex signals (like cosmic strings~\cite{Boileau2022strings} and phase transitions~\cite{banagiri_mapping_2021, Boileau_2023_pt}) for future study and publication. Yet, power laws can be representative of various stochastic source types. A power law with spectral index $n = 2/3$ is usually considered to be a good approximation to describe the \gls{sgwb} from compact binaries~\cite{Regimbau2011rp, LIGOScientific2019vic}, whereas a $n = 0$ power law signal reflects a scale-free cosmological generation mechanism typically driven by early-universe slow-roll inflation scenarios, or by cosmic defect networks~\cite{Bartolo2016ami,Caprini2018mtu} which exhibit scale invariance in the \gls{lisa} band \cite{Caprini2018mtu}. Furthermore, as mentioned in~\cite{Caprini2019pxz} and references therein, spectral indices in the range $0.5 \lesssim  n  \lesssim 1$ in the presence of a kinetic energy-dominated phase (see for example ~\cite{Gouttenoire2021jhk} for a review) can also be considered.

There exists various features that could be exploited in order to test \gls{lisa}'s ability to resolve a \gls{sgwb} signal. The characteristics of the \gls{sgwb} itself, such as its amplitude, the possible particular frequency slope(s) or the possible presence of bumps can be used to distinguish the signal from the noise. The time variability of the \gls{sgwb} for cosmological sources is not expected to provide a useful handle, and for some astrophysical sources, such as Galactic binaries, the effect is expected to be marginal~\cite{Caprini2019pxz}, although accounting for a non-stationary behaviour can help the inference~\cite{Adams2013qma}. One could also try to use anisotropies of the \gls{sgwb} to distinguish different sources as they are characterized by different angular spectra~\cite{LISACosmologyWorkingGroup2022kbp}. 
However, to focus the scope of our study, we will refrain from discussing the possible role of anisotropies. This feature deserves further studies (which could also possibly imply further assumptions on the instrumental noise) and we defer this discussion for future work.

In this work, we take a step towards more realism by using time-domain \gls{lisa} data simulations with time-varying, unequal arms and second-generation time-delay interferometry~\cite{Tinto1999yr,Estabrook2000ef,Tinto2002de,Tinto2003vj,Tinto2014lxa}. As for the data analysis, we introduce flexibility in the noise modelling by fitting generic spline functions to the interferometer noise. While previously used to model the noise \gls{psd} for both LIGO-Virgo~\cite{Littenberg2015, Edwards2015, Chatziioannou2019, edwards_bayesian_2019} and LISA data analysis~\cite{Littenberg2023xpl, Edwards2020}, such a technique has not been tested for \gls{sgwb} detection. We make use of three main sensible features to disentangle \gls{sgwb} from noise: i) a fixed, parametrized signal template; ii) the knowledge of the distinctive transfer functions for noise and \gls{gw} strain and iii) the use of the full covariance matrix of the \gls{tdi} variables. Besides, we rely on two idealizations in this work. First, we assume all non-stochastic \gls{gw} sources have been perfectly subtracted from the data, thus leaving behind idealized residual data. Second, we assume a unique transfer function for the noise. These simplifications allow us to focus on introducing more degrees of freedom in modelling the noise's spectral shape and assess its impact on detection.

The paper is organized as follows. In \cref{sec:simu} we describe the way we simulate the data. In \cref{sec:da-model} we detail the data analysis method including the model assumptions, the likelihood (\cref{sec:likelihood}) and the priors (\cref{sec:priors}) we use. We describe our results on the detection of the \gls{sgwb} signal and the associated parameter estimation in \cref{sec:results} before concluding with a discussion on the results and prospects for future developments in \cref{sec:discussion}.

\section{Data simulation}
\label{sec:simu}

\subsection{Stochastic gravitational-wave background}
\label{sec:sgwb}

A \gls{sgwb} is defined as the superposition of many non-resolvable random signals. Formally, we write the strain as
\begin{equation}
	\label{eq:strain_integral}
	\vb{h}(t) = \int{\vb{h}(t, \vu{n}) \dd{\vu{n}}},
\end{equation}
where we integrate over all possible source directions $\vu{n}$. We use \texttt{LISA GW Response}~\cite{lisagwresponse} to simulate the \gls{sgwb} signal. \texttt{LISA GW Response} approximates this sky integral as a discrete sum over a limited number of point sources $N$ (sky resolution). The stochastic point sources are evenly spread on the celestial sphere using \texttt{HEALPix}\footnote{\url{http://healpix.sourceforge.net}}~\cite{Zonca2019,2005ApJ...622..759G}, with direction vectors $\vu{n}_k$ for $k=1, \dots, N$. The previous equation now reads
\begin{equation}
	\vb{h}(t) = \sum_{k=1}^{N}{\vb{h}(t, \vu{n}_k)}.
\end{equation}
Our model fixes $S_h(f)$, the strain \gls{psd}, defined by the long-duration limit of the expectation of its Fourier transform's square modulus, as
\begin{equation}
	S_h(f, \vu{n}_k) \equiv \lim_{T \to +\infty} \operatorname{E} \left[ \frac{1}{2T} \abs{\int_{-T}^{+T} h_p(t, \vu{n}_k) e^{-2i \pi f t} \dd{t} }^2 \right],
\end{equation}
where we have written the strain in the traceless-transverse gauge for the specific source $k$, hence with the two polarizations $p = +, \times$.

We assume that the spectrum of \gls{gw} energy density per logarithmic frequency intervals at present day is characterized by a power law
\begin{equation}
	\Omega_{\mathrm{GW}}(f) = \Omega_{0} \qty(\frac{f}{f_0})^{n},
	\label{eq:sgwb-energy}
\end{equation}
where $\Omega_{0}$ and $n$ are respectively the energy density at the pivot frequency $f_{0}$ and the spectral index, i.e., the model parameters we will have to estimate. The pivot frequency is chosen at the geometric mean of the bounds of the analysed frequency bandwidth, so that $f_{0} = \sqrt{f_{\mathrm{min}} f_{\mathrm{max}}}$ with $f_{\mathrm{min}} = \SI{0.1}{\milli\hertz}$ and $f_{\mathrm{max}} = \SI{100}{\milli\hertz}$.

We assume that the \gls{sgwb} is isotropic, i.e.,
\begin{equation}
	S_h(f, \vu{n}_k) = \frac{1}{N} S_h(f) \, \forall k,
\end{equation}
and relate $\Omega_{\mathrm{GW}}(f)$ to the one-sided \gls{gw} strain power spectral density as~\cite{Caprini2018mtu}
\begin{equation}
	S_{h}(f) = \Omega_{\mathrm{GW}}(f) \frac{3H_{0}^2}{4 \pi^2 f^3},
	\label{eq:sgwb_psd}
\end{equation}
where $H_{0}$ is the Hubble parameter at present day.

We generate the stochastic point source's strain in the time domain, using an inverse Fourier-transform method.

\subsection{Link response}
\label{sec:link-response}

\begin{figure}
	\centering
	\includegraphics[width=\columnwidth]{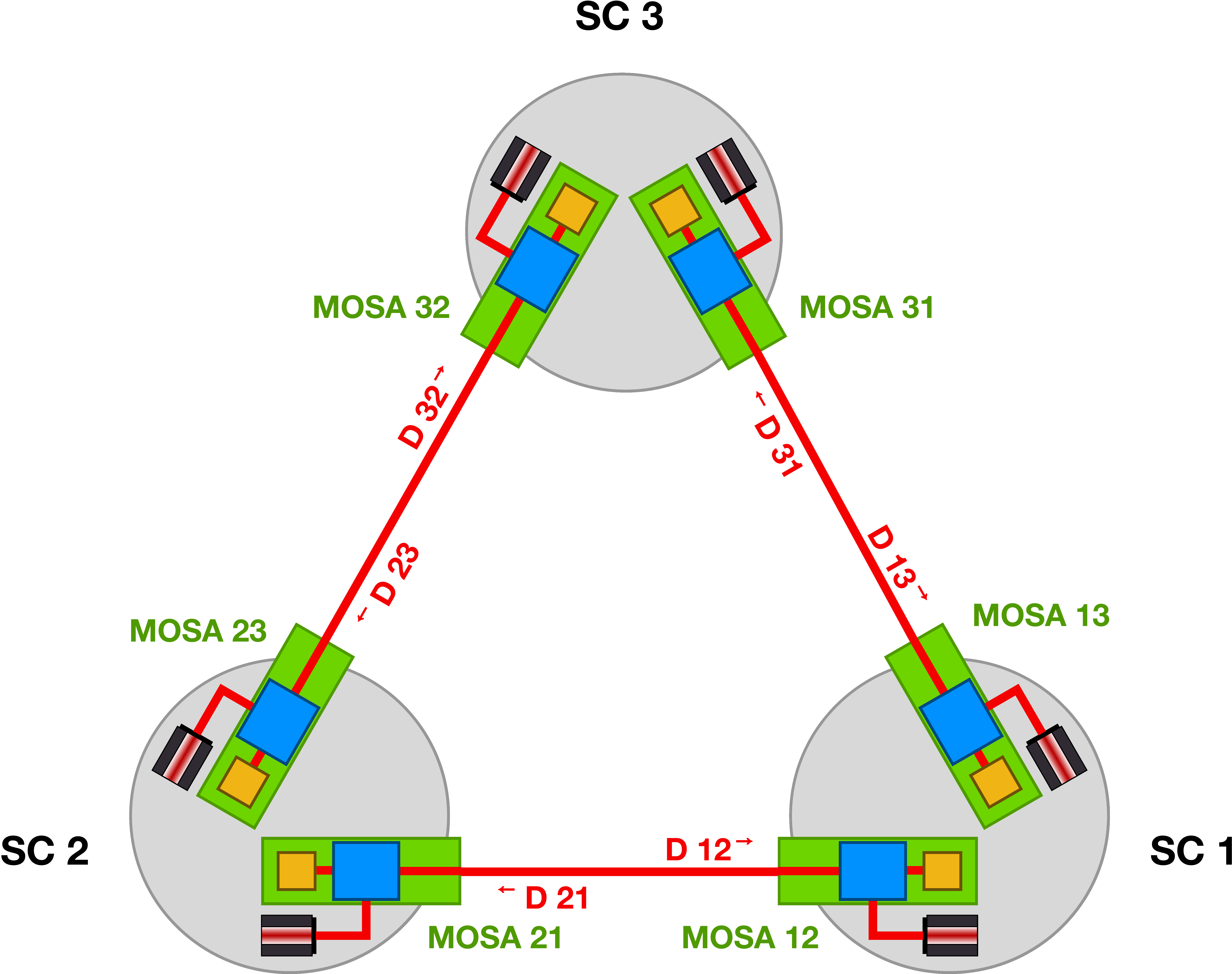}
	\caption{Standard \gls{lisa} conventions. Spacecraft are labelled with 1, 2, 3; \glspl{mosa} are identified with two indices. Elements and quantities uniquely related to one spacecraft or one \gls{mosa} carry the same label.}
	\label{fig:indexing}
\end{figure}

We describe the instrument and the measurements following the standard \gls{lisa} conventions, which are illustrated in \cref{fig:indexing}. Spacecraft are indexed from 1 to 3 clockwise when looking down on the $z$-axis. \Glspl{mosa} are indexed with two numbers $ij$, where $i$ is the index of the spacecraft the system is mounted on (local spacecraft), and $j$ is the index of the spacecraft the light is received from (distant spacecraft).

The \gls{lisa} measurements are labelled according to the \gls{mosa} on which they are performed. Light propagation times are indexed according to the \gls{mosa} on which they are measured, i.e., the receiving \gls{mosa}. In the rest of this paper, we only write quantities for a specific choice of indices (spacecraft or \gls{mosa}), and leave it to the reader to form all remaining expressions using circular permutation and swapping of indices.

The first step to computing the instrument response to the \gls{sgwb} is to compute the deformation induced on the six \gls{lisa} laser links via \texttt{LISA GW Response}. We use the linearity of the response function to write the overall response $y_{12}(t)$ of link $12$ as the discrete sum of the individual link responses to the $N$ point sources,
\begin{equation}
	y_{12}(t) = \sum_{k=1}^N{y_{12,k}(t)}.
	\label{eq:y-sum}
\end{equation}
Similar equations can be written for all 6 \gls{lisa} links.

The time series of frequency shifts $y_{12,k}(t)$, experienced by light traveling along link $12$, is computed by projecting the strain of point source $k$ on the link unit vector (computed from the spacecraft positions). The derivation of the link response, under usual approximations (expansion of the wave propagation time to first order, spacecraft immobile during this propagation time) can be found in \cref{app:derivation-response}, as well as in the literature~\cite[e.g.][]{Cornish2002rt}. It reads
\begin{align}
	y_{12,k}(t) & \approx{}\frac{1}{2 \qty(1 - \vu{k}_k \vdot \vu{n}_{12}(t))}
	\left[ \right. \nonumber \\
	& \left. H_{12,k} \qty(t - \frac{L_{12}(t)}{c}  - \frac{\vu{k}_k \vdot \vb{x}_2(t)}{c})  \right. \nonumber \\
	&\left. - H_{12,k} \qty(t - \frac{\vu{k}_k \vdot \vb{x}_1(t)}{c}) \right].
	\label{eq:instrument-response-to-point-source}
\end{align}
%\begin{align}
%	\begin{split}
	%		y_{12,k}(t) \approx{}& \frac{1}{2 \qty(1 - \vu{k}_k \vdot \vu{n}_{12}(t))}
	%		\\
	%		&\times \left[ H_{12,k} \qty(t - \frac{L_{12}(t)}{c} - \frac{\vu{k}_k \vdot \vb{x}_2(t)}{c}) \right.
	%		\\
	%		&- \left. H_{12,k} \qty(t - \frac{\vu{k}_k \vdot \vb{x}_1(t)}{c}) \right].
	%	\end{split}
%	\label{eq:instrument-response-to-point-source}
%\end{align}

The $y_{ij}$ time series along the 6 \gls{lisa} links are then combined in various ways to compute the \gls{tdi} observables.

\subsection{Instrumental noise}
\label{sec:noises}

We include the dominant secondary noises in our analysis, which are test-mass acceleration noise and readout noise (mainly shot noise). We assume that laser frequency noise is perfectly suppressed by \gls{tdi}, and therefore do not include it in our simulations.

We assume that the noises are uncorrelated in each \gls{mosa}, and identically distributed. The \gls{psd} of test-mass acceleration noise is given by
\begin{align}
	S_{\mathrm{TM}}(f) = a_{\mathrm{TM}}^2 \left[1 + \left(\frac{f_{1}}{f}\right)^2 \right] \left[1 + \left(\frac{f}{f_2}\right)^4\right],
\end{align}
where $a_{\mathrm{TM}} = 3 \times 10^{-15}\,  \mathrm{m s^{-2}}$, $f_{1} = 4 \times 10^{-4}$ Hz and $f_2 = 8$ mHz.
The readout noise \gls{psd} is
\begin{align}
	S_{\mathrm{OMS}}(f)= a_{\mathrm{OMS}}^2 \left[1 + \left(\frac{f_3}{f}\right)^4\right],
\end{align}
where $a_{\mathrm{OMS}} = 15 \times 10^{-12} \, \mathrm{m Hz^{-1/2}}$ and $f_3 = 2$~mHz.

We generate instrumental noise directly at the science interferometer level, assuming no correlations between different interferometers. This way, only the diagonal elements of the links' noise covariance matrix are non-vanishing. While unrealistic, this assumption is meant to simplify the subsequent analysis at relatively small cost in terms of impact on the noise covariance structure (see \cref{sec:tdi}).

\subsection{Time-delay interferometry}
\label{sec:tdi}

\Gls{tdi} combinations are defined as linear combinations of time-shifted measurements. The first and second-generation Michelson combinations, $X_1$ and $X_2$, are given by~\cite{Tinto2003vj},
\begin{align}
	\begin{split}
		X_1 ={}& 
		y_{13} + \delay{13} y_{31} + \delay{131} y_{12} + \delay{1312} y_{21}
		\\
		&- [y_{12} + \delay{12} y_{21} + \delay{121} y_{13} + \delay{1213} y_{31}],
		\label{eq:tdi1}
	\end{split}
	\\
	\begin{split}
		X_2 ={}& X_1 + \delay{13121} y_{12} + \delay{131212} y_{21} + \delay{1312121} y_{13} + \delay{13121213} y_{31}
		\\
		& - [\delay{12131} y_{13} + \delay{121313} y_{31} + \delay{1213131} y_{12} + \delay{12131312} y_{21}],
		\label{eq:tdi2}
	\end{split}
\end{align}
Delay operators are defined by
\begin{equation}
	\label{eq:delay_operators}
	\delay{ij} x(t) = x(t - L_{ij}(t)),
\end{equation}
where $L_{ij}(t)$ is the delay time along link $ij$ at reception time $t$. Because light travel times evolve slowly with time, we compute chained delays as simple sums of delays rather than nested delays, i.e.,
\begin{equation}
	\delay{i_1, i_2, \dots, i_n} x(t) = x\qty(t - \sum_{k=1}^{n-1}{L_{i_k i_{k+1}}(t)} ),
\end{equation}
While this approximation cannot be used to study laser-noise suppression upstream of the \gls{lisa} data analysis, it is sufficient when computing the response function. Note that these equations are left unchanged (up to a sign) by reflection symmetries. However, applying the three rotations generates the three Michelson combinations, $X,Y,Z$, for both generations. In our simulation, we compute them using the \texttt{PyTDI}~\cite{pytdi} software.

\begin{figure}[ht]
	\centering
	\includegraphics[width=\columnwidth]{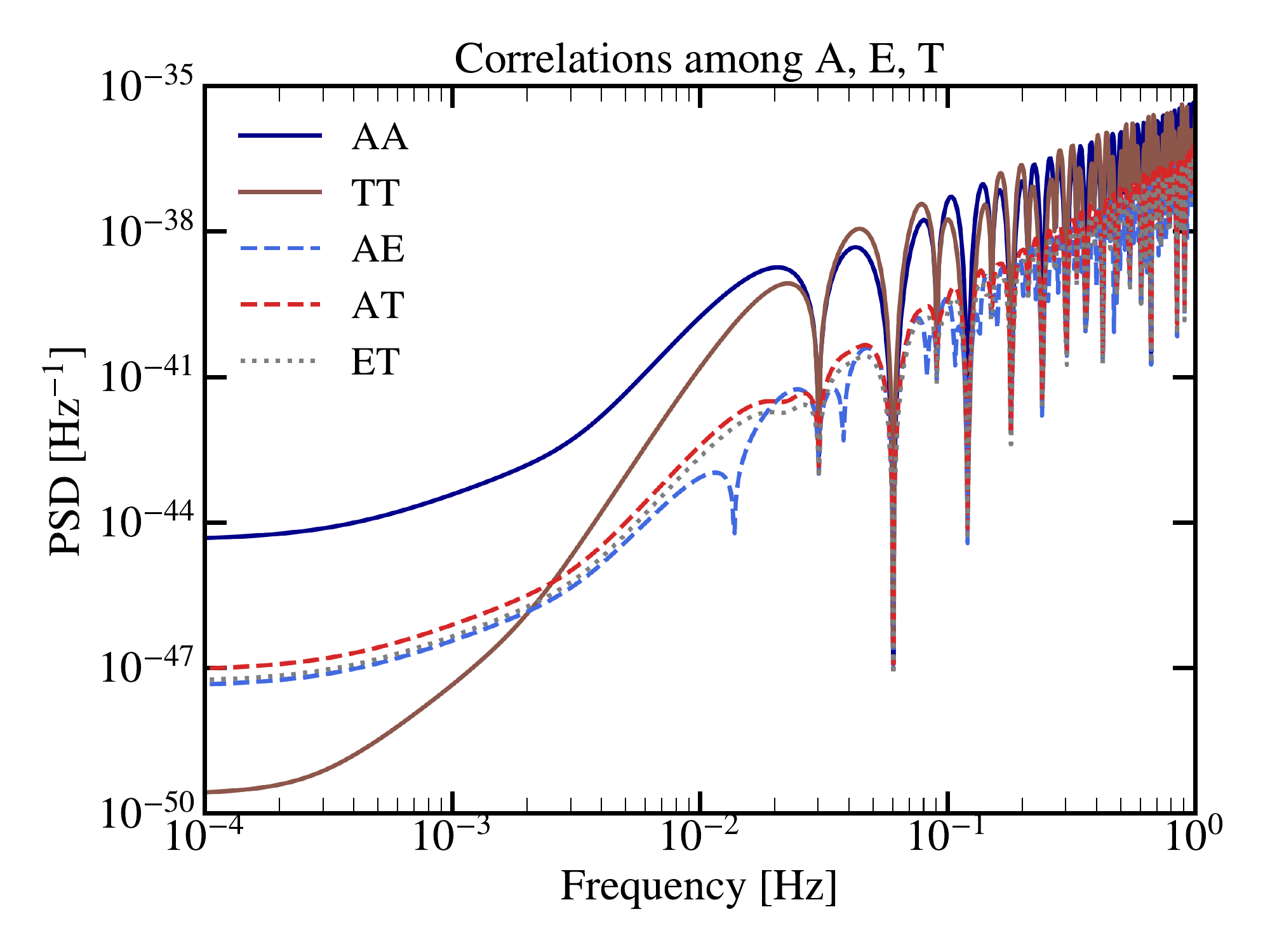}
	\caption{\Glspl{psd} of \gls{tdi} $A = E$ and $T$ (solid dark blue and brown lines, respectively) compared with their \glspl{csd} (light blue, red, and grey dashed lines).}
	\label{fig:aet_correlations}
\end{figure}

Michelson combinations have highly-correlated noises. An quasi-uncorrelated set of \gls{tdi} variables, $A, E, T$, can be obtained from linear combinations of $X,Y,Z$, given by~\cite{Vallisneri2004bn}.
However, $A,E,T$ are only exactly orthogonal (or uncorrelated) under the equal-armlength, equal noise assumptions. In this work, armlengths are not equal, so that we cannot consider $A,E,T$ as exactly uncorrelated. To visualize it, we compute their theoretical \glspl{psd} in Fig.~\ref{fig:aet_correlations}, which shows that below 3 mHz the \gls{csd} levels (dashed curves) become dominant over the $TT$ \gls{psd} (solid brown curve). 
Therefore, we perform the data analysis directly from \gls{tdi} combinations $X,Y,Z$ by modelling their full $3 \times 3$ covariance (see \cref{sec:da-model}). 
\begin{figure}[ht]
	\centering
	\includegraphics[width=\columnwidth]{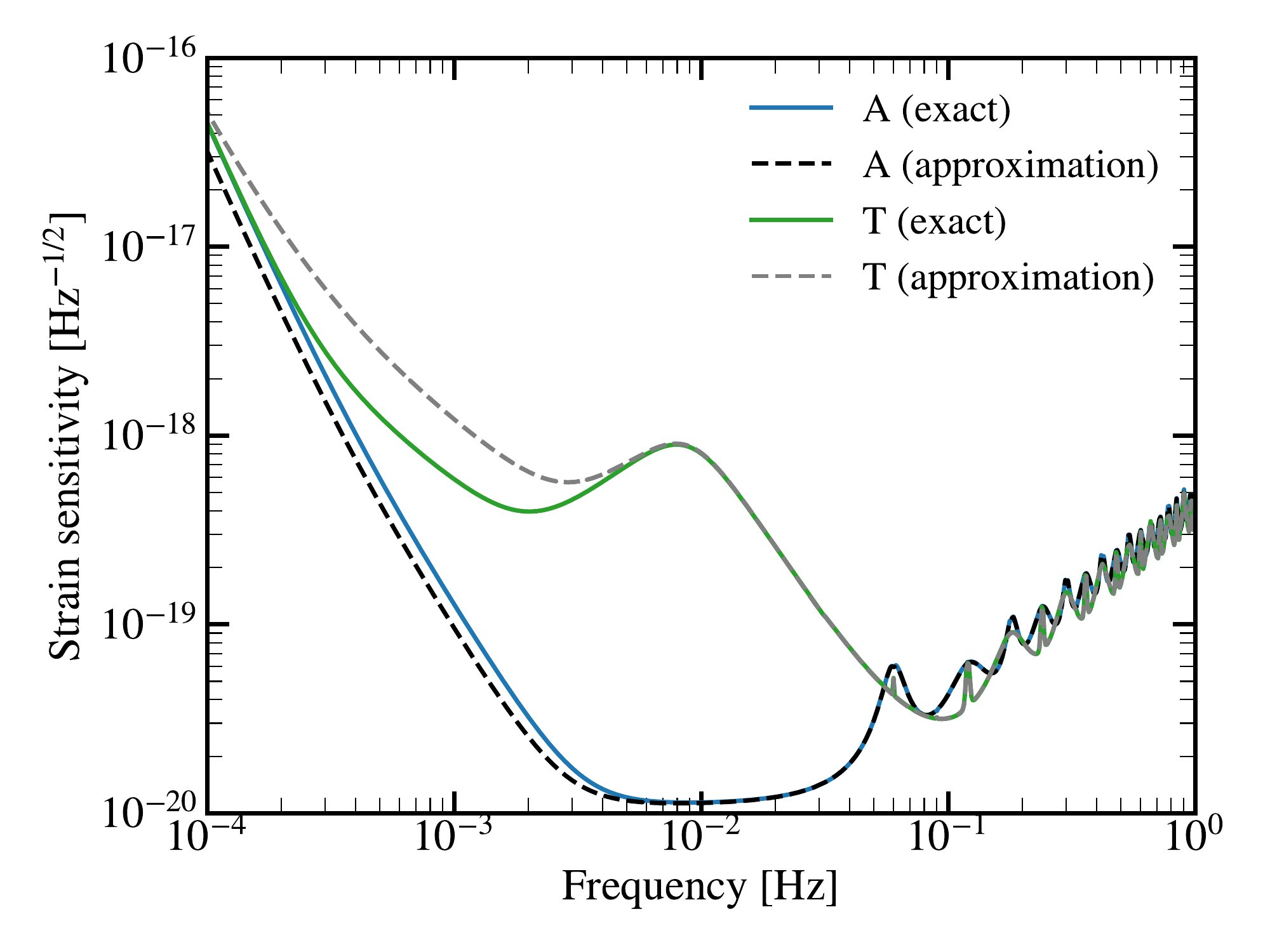}
	\caption{Effect of neglecting the cross-correlations among link measurements on the TDI sensitivity. The \glspl{psd} of \gls{tdi} $A$ and $T$ with correlated links are represented by the continuous blue and green curves, respectively. The effect of assuming uncorrelated links is shown by the dashed black and gray lines. At high frequency (above 5 mHz), the difference is negligible. A deviation appears at low frequency, where the assumption leads to underestimating the strain noise level for both channels.}
	\label{fig:aet_sensitivities}
\end{figure}

We illustrate in \cref{fig:aet_sensitivities} the effect of the assumption we introduced in \cref{sec:noises} when neglecting the cross-correlations among the links $y_{ij}$, where we compare the change in GW sensitivity of \gls{tdi} variables $A$ and $T$ with (solid curves) and without (dashed curves) the uncorrelated link assumption.
While the relative error remains smaller than \SI{5}{\percent} at high frequency, the plot shows a discrepancy of about \SI{50}{\percent} in $A$ and a factor of 4 in $T$ at frequencies below 10 mHz. In other words, the assumption leads to a slight decrease of the overall noise level, and an overestimation of the attenuating power of $T$ at low frequency, which is usually considered as a quasi-null channel. However, the asymptotic behavior is the same: in the low-frequency limit, the $T$ channel is not suppressing gravitational waves better than $A$ or $E$.

\section{Data analysis model}
\label{sec:da-model}

In the analysis, we consider the data vector $\mathbf{\tilde{d}} \equiv (\tilde{X}, \tilde{Y}, \tilde{Z})^T$ of the Fourier-transformed \gls{tdi} variables. For each frequency $f$, we encode the \gls{tdi} transformation of \cref{eq:tdi2} in a matrix $\mathbf{M}_{\mathrm{TDI}}$, so that can write the measured data $\mathbf{\tilde{d}}$ as a function of the link vector $\mathbf{\tilde{y}}$ as %Hence, we can encode the TDI transformation in a $3 \times 6$ matrix $\mathbf{M}_{\mathrm{TDI}}$ applying the right operators to each of the entries of the link vector $\tilde{y}$, defined similarly by
\begin{equation}
	\mathbf{\tilde{d}}(f) = \mathbf{M}_{\mathrm{TDI}}(f) \mathbf{\tilde{y}}(f),
	\label{eq:link2tdi}
\end{equation}
where we defined the link vector as
\begin{equation}
	\mathbf{\tilde{y}} = (\tilde{y}_{12},\, \tilde{y}_{23},\, \tilde{y}_{31},\, \tilde{y}_{13},\, \tilde{y}_{32},\, \tilde{y}_{21})^T.
	\label{eq:y_vector}
\end{equation}
To compute the transfer function $\mathbf{M}_{\mathrm{TDI}}(f)$, it is sufficient to approximate all the delays operators defined in \cref{eq:delay_operators} as complex phasing operators~\cite{Katz:2022yqe},
\begin{equation}
	\delay{ij} \tilde{x}(f) \approx \tilde{x}(f)e^{- 2\pi i f L_{ij}}.
\end{equation}

We assume that the link data is only made of two stochastic processes: the \gls{sgwb} signal $\mathbf{\tilde{y}}_{\mathrm{GW}}$ and the instrumental noise $\mathbf{\tilde{n}}$, so that
\begin{equation}
	\mathbf{\tilde{y}}(f) = \mathbf{\tilde{y}}_{\mathrm{GW}}(f) + \mathbf{\tilde{n}}(f).
\end{equation}
Since signal and noise are independent processes, the \gls{tdi} data covariance can be written as the sum of the \gls{sgwb} and instrumental noise link covariances,
\begin{equation}
	\mathbf{C}_{y}(f) = \langle \mathbf{\tilde{y}} \mathbf{\tilde{y}}^{\dag} \rangle = \mathbf{C}_{\mathrm{GW}}(f) + \mathbf{C}_{n}(f).
\end{equation}
We straightforwardly deduce the \gls{tdi} covariance from \cref{eq:link2tdi} as
\begin{equation}
	\mathbf{C}_{d}(f) = \mathbf{M}_{\mathrm{TDI}}(f) \mathbf{C}_{y}(f) \mathbf{M}_{\mathrm{TDI}}^{\dag}(f).
\end{equation}

Note that it is not necessary to include laser frequency noise, as we assume that it is perfectly canceled by \gls{tdi}. As discussed in \cref{sec:noises}, we further assume that the noises affecting each link measurements are uncorrelated and all characterized by the same one-sided \gls{psd} $S_{n}(f)$. Therefore, their covariance is diagonal:
\begin{equation}
	\mathbf{C}_{n}(f) \equiv \langle \mathbf{\tilde{n}}\mathbf{\tilde{n}}^{\dag} \rangle = \frac{1}{2} S_{n}(f) \mathbf{I}_{6}.
\end{equation}
This assumption allows us to easily express the contribution of the noise to the full covariance as a simple product 
\begin{equation}
	\mathbf{C}_{n}(f) = \frac{1}{2} S_{n}(f) \mathbf{M}_{\mathrm{TDI}}(f) \mathbf{M}_{\mathrm{TDI}}^{\dag}(f).
\end{equation}

As for the \gls{gw} signal, we assume that it is isotropic and stationary, so that its response at any frequency $f$ and time $t_0$ can be encoded in a $6 \times 6$ matrix $\mathbf{R}(f, t_{0})$ as
\begin{equation}
	\mathbf{C}_{\mathrm{GW}}(f) = S_{h}(f) \mathbf{R}(f, t_{0}),
	\label{eq:link-gw-covariance}
\end{equation}
where the elements of $\mathbf{R}(f, t_{0})$ are explicitly derived in \cref{sec:sgwb_response_freq}. The background isotropy brings a quasi-independence on time, so that the choice of $t_0$ is irrelevant in our study.

The key point of the analysis is that we assume that we know both the frequency-dependent \gls{tdi} transfer matrix $\mathbf{M}_{\mathrm{TDI}}(f)$ and the \gls{gw} response matrix $\mathbf{R}(f, t_0)$. Both of them depend on inter-spacecraft distances, which we suppose we know perfectly. Then, the parameters we have to estimate are the ones describing the signal \gls{psd} $S_{h}(f)$ and the noise \gls{psd} $S_{n}(f)$. 
\Cref{eq:sgwb-energy} provides the parametrization of $S_{h}(f)$, which includes the energy density $\Omega_{0}$ and spectral index $n$. The model for $S_{n}(f)$ is detailed in the next section.

\subsection{Noise model}

We aim at having a generic and flexible modeling of the noise. To this end, we model the single-link noise log-\gls{psd} with interpolating cubic B-spline functions. This basis provides a stable parametrization of any sufficiently smooth function, avoiding numerical errors that can arise when using high order polynomials. The parameters of the model are the logarithm of the control frequencies $x_i$ and their corresponding log-\gls{psd} ordinates $a_{i}$. We fix the first and last control frequencies to be the boundaries of the analysed frequency bandwidth, so that $x_{0} = \log f_{\min}$ and $x_{Q} = \log f_{\max}$, where $Q+1$ is the total number of control points. Then, we construct the spline function
\begin{equation}
	\log S_{n}(f) = \sum_{i=1}^{Q+1} a_{i}  B_{i,3}\left(\mathbf{\xi}, f \right),
\end{equation}
where $a_i$ are the spline coefficients and $\mathbf{\xi}$ is the vector of the $Q+5$ spline knots. The basis elements $B_{i,3}(\mathbf{\xi}, f)$ are defined recursively as
\begin{align}
	B_{i, 0}(f) & = 1, \textrm{ if $\xi_i \le \log f < \xi_{i+1}$, otherwise $0$,} \nonumber \\
	B_{i, k}(f) & = \frac{\log{f} - \xi_i}{\xi_{i+k} - \xi_i} B_{i, k-1}(f) + \frac{\xi_{i+k+1} - \log{f}}{\xi_{i+k+1} - \xi_{i+1}} B_{i+1, k-1}(f).
\end{align}
The spline knots are directly related to the control points as
\begin{equation}
	\begin{split}
		\xi_{i} & = x_{0}\; \forall i \in [0, \, 3];  \\
		\xi_{i+ 3} & = x_{i} \; \forall i \in [0, \, Q];  \\
		\xi_{i + 3} & = x_{Q}  \; \forall i \in [Q, \, Q+3];  \\
		\log S_{n}(e^{x_{i+3}}) & = a_{i} \; \forall i \in [0, \, Q].
	\end{split}
\end{equation}

In practice, we use the \textsc{interp1d} function of the \textsc{SciPy} package~\cite{virtanen_scipy_2020}, which builds the spline basis based on the control log-frequencies $x_{i}$ and their corresponding ordinates $a_i$. Since the frequencies of the first and last control points are fixed, the spline model is described by $2Q$ parameters that we can gather in a vector $\boldsymbol{\theta}_{n} = (x_{0}, \dots, x_{Q}, a_{1}, \dots, a_{Q-1})^T$.

\subsection{Likelihood}
\label{sec:likelihood}

In principle, one could directly write down the likelihood for the frequency-domain \gls{tdi} data $\mathbf{\tilde{d}}$ using Whittle's approximation~\cite{Whittle1953}. To decrease the computational cost of the likelihood evaluation, we instead consider frequency sample averages of the periodogram.

Let us define the normalized windowed \gls{dft} of any multivariate time series of length $N_{x} = \lfloor T f_{s} \rfloor$ as
\begin{equation}
	\mathbf{\tilde{x}}(f_k) = \sqrt{\frac{2}{\kappa_{2} f_{s}}} \sum_{n=0}^{N_{x}-1} w_{n} \mathbf{x}_{n} e^{-2\pi k n / N_{x}},
	\label{eq:dft}
\end{equation}
where $w_{n}$ is a time window smoothly decreasing to zero at the edges of the time series, and $\kappa_{p} = \sum_{n=0}^{N_{x}-1} {w_n}^p$. We choose this normalization such that the periodogram is directly given by the square modulus of $\tilde{\mathbf{x}}_{k}$, and its expectation is directly comparable with the one-sided \gls{psd}.

At each frequency bin $f_k$, we define the $3 \times 3$ periodogram matrix as $\mathbf{P}(f_k) \equiv \mathbf{\tilde{d}}(f_k) \mathbf{\tilde{d}}(f_k)^{\dag}$.
To compress the data, we split the frequency series $\mathbf{\tilde{d}}$ into $J$ consecutive, non-overlapping segments. We call $f_j$ the central frequency and $n_{j}$ the size of each segment $j$. We define the averaged periodogram matrix $\mathbf{\bar{P}}(f_j)$ by averaging the periodograms over the frequency bins within each segment $j$,
\begin{equation}
	\mathbf{\bar{P}}(f_{j}) \equiv \frac{1}{n_{j}} \sum_{k=j-\frac{n_j}{2}}^{j+\frac{n_j}{2}} \mathbf{\tilde{d}}(f_k) \mathbf{\tilde{d}}(f_k)^{\dag}.
	\label{eq:averaged_periodogram}
\end{equation}
If the \glspl{dft} $\mathbf{\tilde{d}}(f_k)$ were uncorrelated between different frequency bins, the matrix $\mathbf{Y}(f_j) \equiv \nu(f_j) \mathbf{P}(f_{j})$ would follow a complex Wishart distribution with $\nu(f_j) = n_j$ \glspl{dof} and scale matrix $\mathbf{C}_{d}(f)$, with a probability density function
\begin{equation}
	p({\mathbf{Y}(f)} \vert \boldsymbol{\theta} )= \frac {\qty|\mathbf{Y}(f)|^{\nu-3} \operatorname{exp}\qty[-\operatorname{tr}(\mathbf {C}_{d}^{-1}\mathbf{Y}(f) )]}{\qty|\mathbf{C}_{d}(f)|^{\nu} \cdot \mathcal{C} \tilde \Gamma_3 (\nu)}, 
	\label{eq:wishart_distribution}
\end{equation}
where $\mathcal{C} \tilde \Gamma_p(\nu)$ is the complex gamma function, $\operatorname{tr\left(\cdot\right)}$ is the trace operator and $\qty|\mathbf{A}|$ is the determinant of any matrix $\mathbf{A}$.
%${\mathcal {C}}{\widetilde {\Gamma }}_{p}^{}(n) \equiv \pi ^{p(p-1)/2}\prod _{j=1}^{p}\Gamma (n-j+1)$ 
In reality, the frequency bins that are close to each other are mildly correlated, depending on the choice of the window function $w_n$ in \cref{eq:dft}. As a result, the effective number of \glspl{dof} $\nu(f_j)$ is smaller than the number of averaged frequency bins $n_j$. A good measure of the reduction factor is provided by the normalized equivalent noise bandwidth $N_{\mathrm{bw}}$, defined for any window $w$ and time series size $N_{d}$ as
\begin{equation}
	N_{\mathrm{bw}} = N_{d} \frac{\kappa_{2}}{\kappa_{1}},
\end{equation}
which is expressed in number of frequency bins. Values of $N_{\mathrm{bw}}$ for various windows can be found in~\cite{heinzel_spectrum_2002}. The effective number of \glspl{dof} is then given by $\nu(f_j) = n_{j} / N_{\mathrm{bw}}$.

Taking the logarithm of \cref{eq:wishart_distribution} above and keeping only the terms depending on the parameters $\boldsymbol{\theta}$ yields
\begin{equation}
	\log p({\mathbf{Y}(f)} \vert \boldsymbol{\theta} )= -\operatorname {tr} (\mathbf {C}_{d}^{-1}\mathbf {Y}(f) ) - \nu(f) \log \qty|\mathbf {C}_{d}(f)|.
	\label{eq:log-likelihood}
\end{equation}
The full log-likelihood across the analyzed bandwidth is then the sum over all frequency bins
\begin{equation}
	%\log p({\mathbf{\bar{P}}} \vert \boldsymbol{\theta} )
	\mathcal{L}_{\mathbf{Y}}(\boldsymbol{\theta}) = \sum_{j=0}^{J-1} \log p({\mathbf{Y}(f_j)} \vert \boldsymbol{\theta} ).
\end{equation}
When both noise and signal are included in the likelihood, the vector of model parameters $\boldsymbol{\theta}$ includes the control point locations, the spline coefficients, and the \gls{gw} parameters $\boldsymbol{\theta} = (x_{0}, \hdots, x_{Q}, a_{1}, \hdots, a_{Q-1}, \log \Omega_{0}, n)^{T}$.

\subsection{Priors}
\label{sec:priors}

Aiming at a robust analysis, we choose poorly constraining priors for the noise parameters. We let the control points take value in an interval bounded by one order of magnitude below and above the true noise model $S_{n, \mathrm{true}}$ (which is used for the injection). This way, we have
\begin{equation}
	S_{n}(f) \in \left[10^{-1}\, S_{n, \mathrm{true}}(f) ;\;  10 \, S_{n, \mathrm{true}}(f)\right].
\end{equation}
Note that this prior does not reflect the allocated margins for the required \gls{lisa} sensitivity, but enables us to remain conservative in our analysis.

We allow the control frequencies to take value within the analyzed bandwidth $\left[ f_{\mathrm{min}}, f_{\mathrm{max}} \right]$. To enforce a relatively even distribution of the control points, we assign to each of them a Beta distribution conditioned on the location of the previous one, such that
\begin{eqnarray}
	p\left( x_{i} \vert x_{i-1} \right) \propto u_{i}^{\alpha_{i} -1}(1-u_i)^{\beta_{i} -1},
\end{eqnarray}
where $u_{i} \equiv \left(x_{i} - x_{i-1} \right) / \left( x_{Q}- x_{i-1} \right)$ is the position of control point $x_i$ relative to the previous one $x_{i-1}$, rescaled in the interval $[0,\, 1]$. We choose parameters values $\alpha_i = 2$ and $\beta_i = Q - i + 2$ so that the mode of the conditional distribution peaks at $( x_{Q} - x_{i-1})/(Q-i)$. This choice ensures that if the control point $x_{i-1}$ is given, as there are $Q-i$ control points left to be placed, the next one has more probability to be placed in the first $1/(Q-i)\mathrm{^{th}}$ of the remaining frequency band.

Concerning the \gls{sgwb} parameters, we impose uniform priors on $\log \Omega_{0}$ and on $n$, respectively in intervals $[-35, \, -28]$ and $[-5, 7]$.

\section{Detection and parameter estimation}
\label{sec:results}

\subsection{Detection}
\label{sec:detection}

In a Bayesian framework, detecting the presence of a stochastic process can be done through model comparison: one model assumes that the data only contains noise (null hypothesis $H_0$), while the other model assumes the presence of a \gls{sgwb} in addition to the noise (tested hypothesis $H_1$). We compare the models by computing their Bayes factor, defined as the ratio of their evidences. The log-Bayes factor is then
\begin{equation}
	\log \mathcal{B}_{10}(\mathbf{Y}) = \log Z_{1}(\mathbf{Y}) - \log Z_{0}(\mathbf{Y}),
	\label{eq:bayes_fact}
\end{equation}
where $Z_i(y) \equiv \int_{{\boldsymbol{\theta}}_{i}} p\left(y \vert H_i\right) d \boldsymbol{\theta}$ is the evidence of the model under hypothesis $H_i$ and $\Theta_{i}$ is the space in which $\boldsymbol{\theta}_{i}$ is allowed to take values. The presence of a \gls{sgwb} is claimed when the Bayes factor stands above a given threshold.

When dealing with parallel-tempered \gls{mcmc} outputs, we can approximate the evidence by thermodynamic integration~\cite{lartillot_computing_2006},
\begin{equation}
	\log Z_{i}(\mathbf{Y}) =\int_{0}^{1} \mathrm{E}_{\beta}\qty[\log p\left(\mathbf{Y} \vert \boldsymbol{\theta}, H_{i} \right)] \dd{\beta},
\end{equation}
where the expectation $\mathrm{E}_{\beta}$ is taken with respect to the tempered posterior density $p_{\beta}\left(\mathbf{Y} \vert \boldsymbol{\theta}, H_{i} \right) \propto  p\left(\mathbf{Y} \vert \boldsymbol{\theta}, H_{i} \right)^\beta p(\boldsymbol{\theta}, H_i)$. The variable $\beta$ is the inverse temperature of the tempered chain, and $\mathrm{E}_{\beta}$ is the expectation of the chain at temperature $1/\beta$ taken over the parameter space $\Theta$. 

\subsection{Averaged Bayes factors}
\label{sec:avg-bayes}

We aim to find the parameter pairs $(\Omega_0, n)$ for which the Bayes factor is equal to the detection threshold $\mathcal{B}_{\mathrm{thresh}}$. To do that, we compute the posterior distributions under both $H_0$ and $H_1$ for a wide range of parameter values. The Bayes factor depends on the specific data realization; instead of generating hundreds of data realizations for each parameter pair, we choose to consider the averaged Bayes factor, that we define as the Bayes factor computed from the expected likelihood under the true distribution when $H_1$ is true.

In other words, if the data is described by the true parameter vector $\boldsymbol{\theta}^{\star}$, then we can compute the averaged log-Bayes factor 
\begin{equation}
	\overline{\log \mathcal{B}_{10}} = \log \mathcal{B}_{10}(\mathbf{\bar{Y}}),
\end{equation}
where $\mathbf{\bar{Y}} = \operatorname{E}_{\theta^{\star}}[\mathbf{Y}]$ is the expectation of the data $\mathbf{Y}$ under the true hypothesis. Note that $\overline{\log \mathcal{B}_{10}}$ is not the statistical expectation of the log-Bayes factor, but we will show later that using $\overline{\log \mathcal{B}_{10}} = \mathcal{B}_{\mathrm{thresh}}$ provides a conservative criterion for detection.

\subsection{Optimal model order}
\label{sec:modelorder}

For this work, we adopted a spline model that is flexible enough to fit the spectral series. With the right parametrization, it yields satisfactory results in inferring the instrumental noise \gls{psd} shape (see \cref{sec:detectability}). One of the challenges of this strategy is to choose the most suitable model order, i.e., the optimal number of spline knots. This is crucial for avoiding over-fitting situations, but also biases in the search and in parameter estimation.

As described previously, we perform a model selection by computing Bayes factors between two hypotheses. Thus, for a given data scenario, we can either perform the analysis multiple times with different spline orders, or dynamically estimate the model order together with its corresponding parameters. As a cross-validation test for our analyses here, we choose the latter applied on a simplified case. We use a \gls{rj}-\gls{mcmc} algorithm~\cite{rjmcmc1}, which is generalization of the Metropolis-Hastings~\cite{Metropolis1959, Hastings1970, MARTINO2018134} algorithm, capable of searching in parameter spaces of varying dimensionality (see~\cite{Christensen2022PE} for a review of sampling techniques). In particular, we use a \gls{rj} algorithm enhanced with parallel tempering techniques~\cite{Vousden2016, emcee, karnesis-in-prep} to efficiently identify the optimal number of knots in our spline model.

To simplify the procedure, we focus on instrumental noise only. We simulate one year of noise data, as described in \cref{sec:simu}, without any \gls{gw} signal present. We then build a likelihood function that is computationally efficient.

\begin{figure*}	
	\centering
	\begin{tabular}{ll}
		\includegraphics[width=.8\columnwidth]{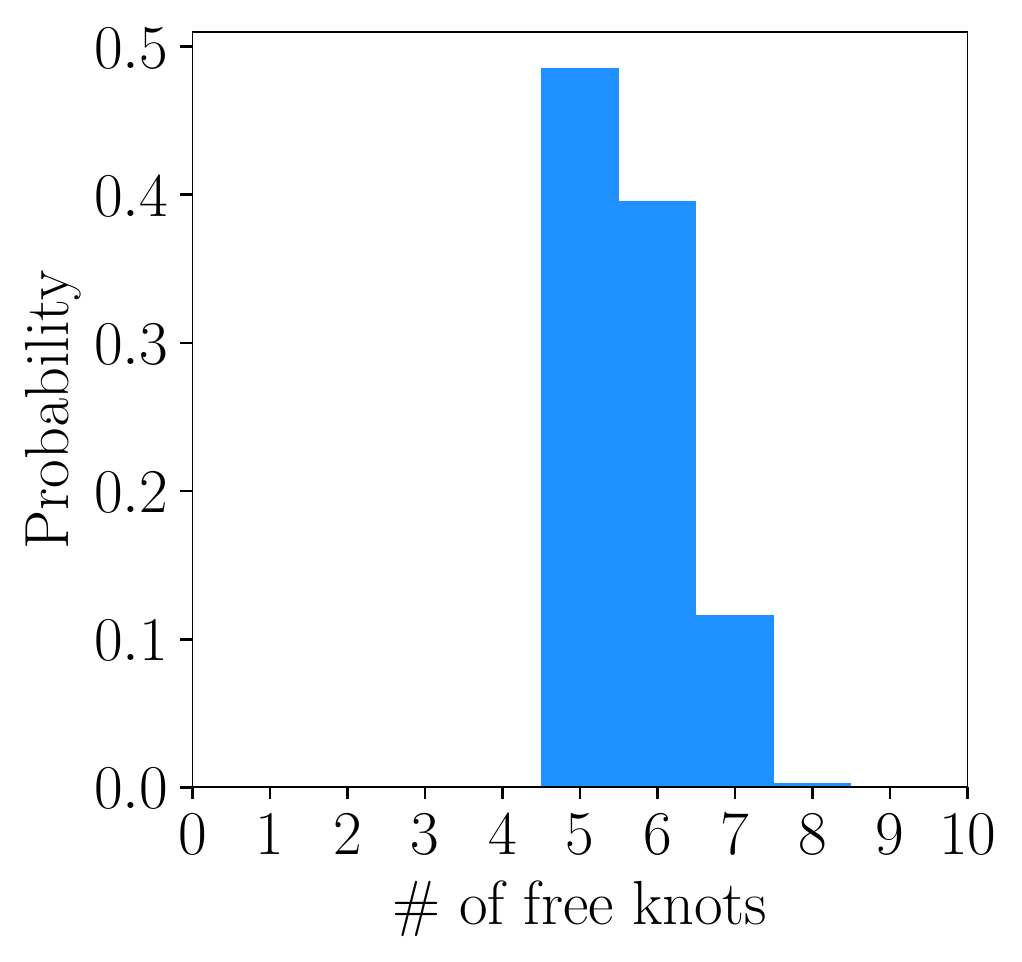} &   
		\includegraphics[width=.8\columnwidth]{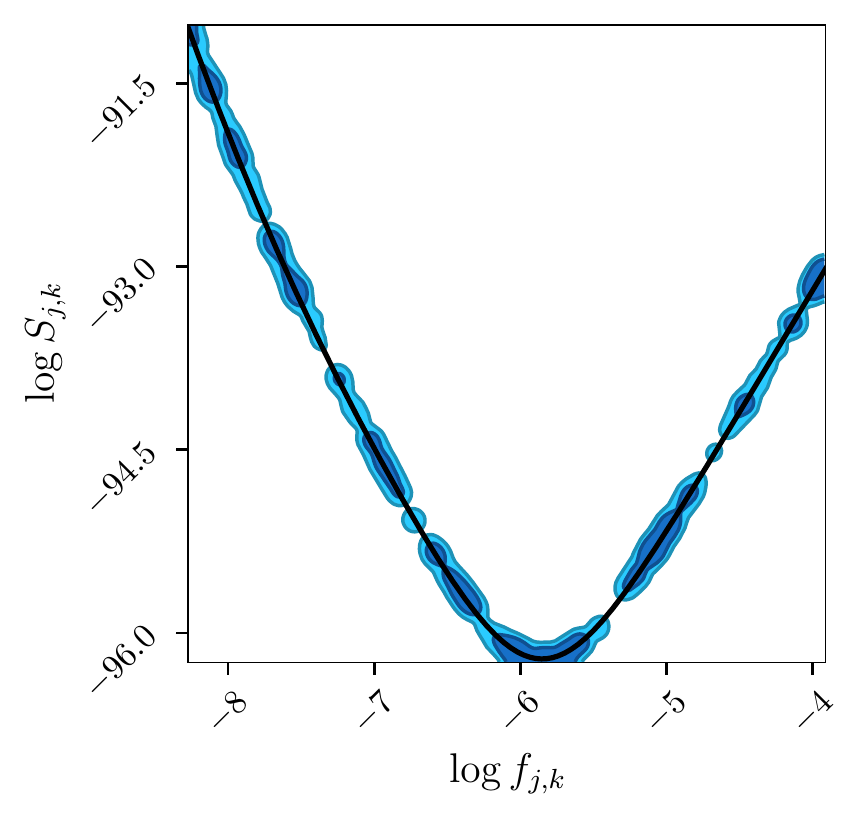}
	\end{tabular}
	\caption{Estimating the optimal model order using \gls{rj}-\gls{mcmc}. In this investigation, we have fixed the frequencies of the two edge knots, while letting the algorithm determine the optimal number of internal knots, together with their frequencies and amplitude. Right: posterior samples of the knots amplitudes $\log S_{j,k}$ and frequencies $\log f_{j,k}$ for {\it all} given spline models of order $k$, as sampled with our \gls{rj} algorithm (we stack the chains for all $k$). The algorithm explores the true noise curve (solid black line) by proposing spline knots positioned across the frequency range (see main text for more details). The plot was generated with~\cite{chainconsumer}.}
	\label{fig:optimal_model_order}
\end{figure*}

Our spline model fixes the control frequencies of the two knots at the edges of our spectrum; their amplitudes $S_\mathrm{low}$ and $S_\mathrm{high}$ are left as free parameters to be estimated. The number of other knots $k$, their frequencies $S_{j,k}$ and amplitudes $f_{j,k}$ in-between are also determined from the data. We remind here that the $j$ index corresponds to the spline number for the given model order $k$.

For the knot parameters, we have chosen a quite broad uniform prior, $\log S_{j,k} \sim \mathcal{U} [-100,\, -91]$ and for $\log f_{j,k}$, a uniform prior across the log-frequency range; for the spline model order $k$, we used an uninformative prior $k \sim \mathcal{U} [3,\, 30]$. Running the algorithm for 10 temperatures~\cite{Vousden2016} with 20 walkers each~\cite{emcee} yields the result shown in the left panel of \cref{fig:optimal_model_order}.

It is particularly interesting to also inspect the 2D posterior slices of the parameters, shown in the right panel of \cref{fig:optimal_model_order}. We have essentially sampled the full parameter space of $\log S_{j,k}$ and $\log f_{j,k}$ for all the  possible values of the dimensionality $k$ of the model. The figure shows that there is no unique solution when fitting both the frequencies and amplitudes of the spline knots, and the \gls{mcmc} chains explore the true shape of the noise spectra.

From the posterior distribution of the model order $k$, shown in the left panel of \cref{fig:optimal_model_order}, we see that a maximum can be found between $k=5$ and $k=6$. In the rest of the study, we fix the model order to this optimal value $k=5$, i.e. $5+2$ knots. This translates to twelve parameters (the internal knots' frequencies and amplitudes, plus the frequencies of the two edge knots). 

\subsection{Assessment of the detectability of a stochastic gravitational-wave background}
\label{sec:detectability}

\begin{figure}[ht]
	\centering
	%<left> <lower> <right> <upper>
	\includegraphics[width=\columnwidth, trim={0.6cm 0.8cm 1cm 0.6cm}, clip]{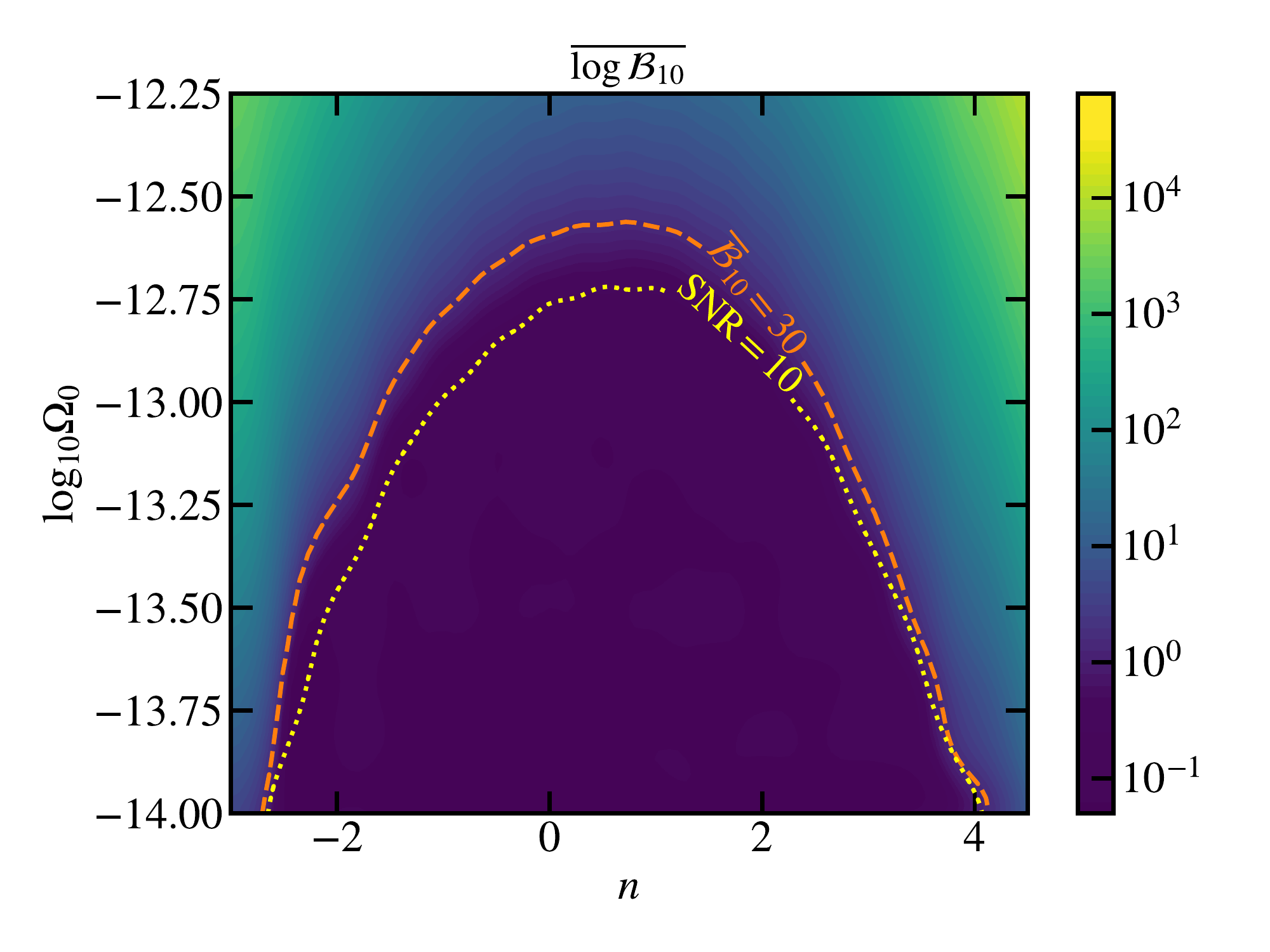} 
	\caption{Left: averaged log-Bayes factor contour plot for a range of \gls{sgwb} spectral index $n$ ($x$-axis) and log-energy density $\log \Omega_{0}$ ($y$-axis) with a knee frequency of $f_{0} = \SI{3.16}{\milli\hertz}$. The color map represents the values of the decimal logarithm of Bayes factor, with warmer shades indicating larger values. The orange dashed line is the detection threshold taken equal to 30, considered as a strong evidence for the presence of the \gls{sgwb}. The yellow dotted line shows the \gls{snr}-10 line as a comparison.}
	\label{fig:bayes_factor_contours}
\end{figure}

Now we compute the averaged Bayes factors for a wide range of \gls{sgwb} parameters to assess our ability to detect a \gls{sgwb} with a noise of unknown spectral shape, under the assumptions that we stated in \cref{sec:da-model}. For a set of spectral indices ranging from -4 to 5, and log-energy densities between \num{E-14} and \num{E-12}, we run our Bayesian model comparison and plot the results in \cref{fig:bayes_factor_contours}. 

We represent values of log-Bayes factors using a color scale, with warmer colors signify large detection evidences. From the initial set of 272 computed point, we interpolate the log-Bayes factor values on a finer grid of $100 \times 100$ points using a Gaussian process regression. This allows us to plot a line of constant Bayes factor (dashed orange) of $\overline{\mathcal{B}_{10}} = 30$, which is considered as a detection threshold for strong evidence for hypothesis $H_1$~\cite{Adams2010vc}. All couples of parameters that lie below this line are considered as undetectable signals, and all above values are strong detections. For example, we find that the amplitude detection threshold for a scale-invariant \gls{sgwb} ($n = 0$) is about $\Omega_{0} = \num{2.5E-13}$, which is close to what previous work using a parametrized noise \glspl{psd} model found (for example, Adams and Cornish get $\Omega_{0} = \num{1.7E-13}$). Besides the obvious effect of the increase of detectability with the energy density, we also observe a dependence that is strongly tied to the spectral shape of the noise present in the data. For a given energy density, the Bayes factor is minimum when $n$ is between 0.5 and 1. We observe the same minimum for the \gls{snr} curve, suggesting that our ability to detect the signal is mainly driven by its \gls{snr}, which is itself determined by both $\Omega_{0}$ and $n$.

The location of the \gls{snr} minimum is set by the strain sensitivity curve in \cref{fig:aet_sensitivities}, as well as the \gls{sgwb} strain \gls{psd}'s dependence on frequency, which is proportional to $f^{n-3}$, as shown in \cref{eq:sgwb_psd}. Note that the choice of the knee frequency (of about \SI{3}{\milli\hertz}) also drives the location of the minimum through its contribution to the effective \gls{sgwb} amplitude.

\Cref{fig:bayes_factor_contours} provides us with the range of power-law parameters that \gls{lisa} will be able to probe. This result can be considered in the context of previous measurements. The LIGO, Virgo and KAGRA collaborations are able to put upper limits on the isotropic gravitational-wave background from Advanced LIGO's and Advanced Virgo's third observing run~\cite{KAGRA:2021kbb}. In particular, they find that the dimensionless energy density is bounded as $\Omega_{\mathrm{GW}} \leq \num{5.8E-9}$ at the \SI{95}{\percent} credible level for a frequency-independent gravitational-wave background, 
%using a prior which is uniform in the log of the strength of the gravitational wave background, 
with \SI{99}{\percent} of the sensitivity coming from the band \SIrange{20}{76.6}{\hertz}. They also find the upper limit $\Omega_{\mathrm{GW}} \leq \num{3.4E-9}$ at \SI{25}{\hertz} for a power-law gravitational-wave background with a spectral index of 2/3 in the band \SIrange{20}{90.6}{\hertz}, and $\Omega_{\mathrm{GW}} \leq \num{3.9E-10}$ at \SI{25}{\hertz} for a spectral index of 3, in the band \SIrange{20}{291.6}{\hertz}.

The NANOGrav collaboration~\cite{NANOGrav2020bcs}, using their \SI{12.5}{\year} pulsar-timing data set, finds that under their fiducial model, the Bayesian posterior of the amplitude has median $1.92^{+0.75}_{-0.55} \times 10^{-15}$ 
%and \SI{5}{\percent} - \SI{95}{\percent} quantiles of $1.37 - 2.67 \times 10^{-15}$  
for an $f^{-2/3}$ spectrum (as expected from a population of inspiralling supermassive black holes) at a reference frequency of \SI{1}{\per\year}. The \gls{ipta} collaboration~\cite{Antoniadis:2022pcn}, using their second data release and for a spectral index of $-2/3$, finds a recovered amplitude of $2.8^{+1.2}_{-0.8} \times 10^{-15}$ at a reference frequency of \SI{1}{\per\year}.

We gather these experimental measurements in \cref{tab:constraints} and compare them to what \gls{lisa} could observe, should the frequency dependence of the \gls{gw} background remain constant in-between the detectors sensitive bands. This comparison shows that \gls{lisa} would be able to detect, or place tighter constraints, on energy densities for \gls{sgwb} searched in LIGO-Virgo or pulsar timing array data. Besides, the detection limits of about \num{E-14} we obtain in \cref{fig:bayes_factor_contours} for extreme spectral indices like $n=-3$ or $n=4$ would yield huge amplitudes in the \gls{ipta} and LIGO-Virgo bands, respectively. Those lying well above the detectors sensitivity, such power laws would be visible today and are therefore not expected to arise in \gls{lisa}.

\begin{table}
	\centering 
	\begin{tabular}{cccccc}
		\toprule
		Detector & $n$ & $\Omega_{\mathrm{det}}(f_{\mathrm{det}})$ & $\Omega_{\mathrm{LISA}}(f_0)$ & Thresh. & Refs \\
		\midrule
		LVK  &   $0$  & $5.8 \cdot 10^{-9}$ &  $5.8 \cdot 10^{-9}$ & $2.5 \cdot 10^{-13}$ & \cite{KAGRA:2021kbb}  \\
		LVK & $2/3$ & $3.4 \cdot 10^{-9}$ & $8.3 \cdot 10^{-12}$ & $2.7 \cdot 10^{-13}$ & \cite{KAGRA:2021kbb} \\
		NANOGrav & $-2/3$ & $1.9 \cdot 10^{-15}$ & $9.2 \cdot 10^{-9}$ & $2.0 \cdot 10^{-13}$ & \cite{NANOGrav2020bcs} \\
		\gls{ipta} & $-2/3$ & $2.8 \cdot 10^{-15}$ & $1.3 \cdot 10^{-8}$ & $2.0 \cdot 10^{-13}$ & \cite{Antoniadis:2022pcn} \\
		\bottomrule
	\end{tabular}
	\caption{Comparison of \gls{lisa} capabilities with current detector constraints on \gls{sgwb} parameters. The columns from left to right show, respectively, the detector's collaboration name; the power-law index value; the energy density computed at the detector pivot frequency $f_\text{det}$ (\SI{25}{\hertz} for LVK, \SI{1}{\per\year} for NANOGrav and \gls{ipta}); the extrapolated energy density at \gls{lisa}'s \SI{3}{\milli\hertz} pivot frequency; the detection threshold computed in this study; and the reference from which we extract the constraints.}
	\label{tab:constraints}
\end{table}

\subsection{Parameter estimation}
\label{sec:pe}

\begin{figure}[ht]
	\centering
	\includegraphics[width=\columnwidth, trim={0 0.5cm 0 0.5cm}, clip]{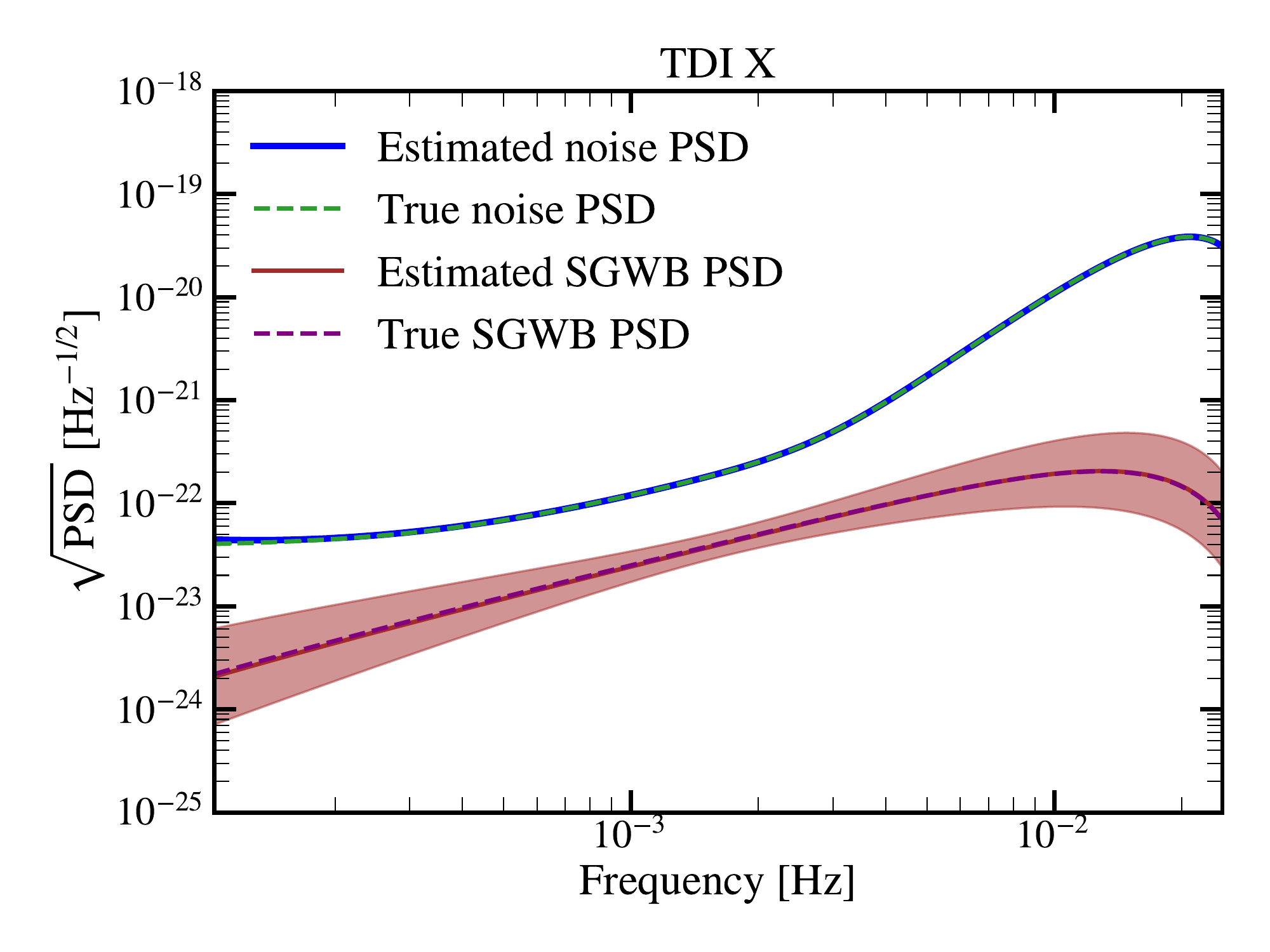}
	\caption{Posteriors of the noise (blue) and the \gls{sgwb} (red) \glspl{psd} in \gls{tdi} channel $X$ for an injection with $\Omega_{0} = \num{1E-13}$ and $n=-1$. The light red-shaded area represents the 3-$\sigma$ credible interval.}
	\label{fig:posterior_1e-13_n-1}
\end{figure}

\begin{figure}[ht]
	\centering
	\includegraphics[width=\columnwidth, trim={0cm 0cm 0cm 0cm}, clip]{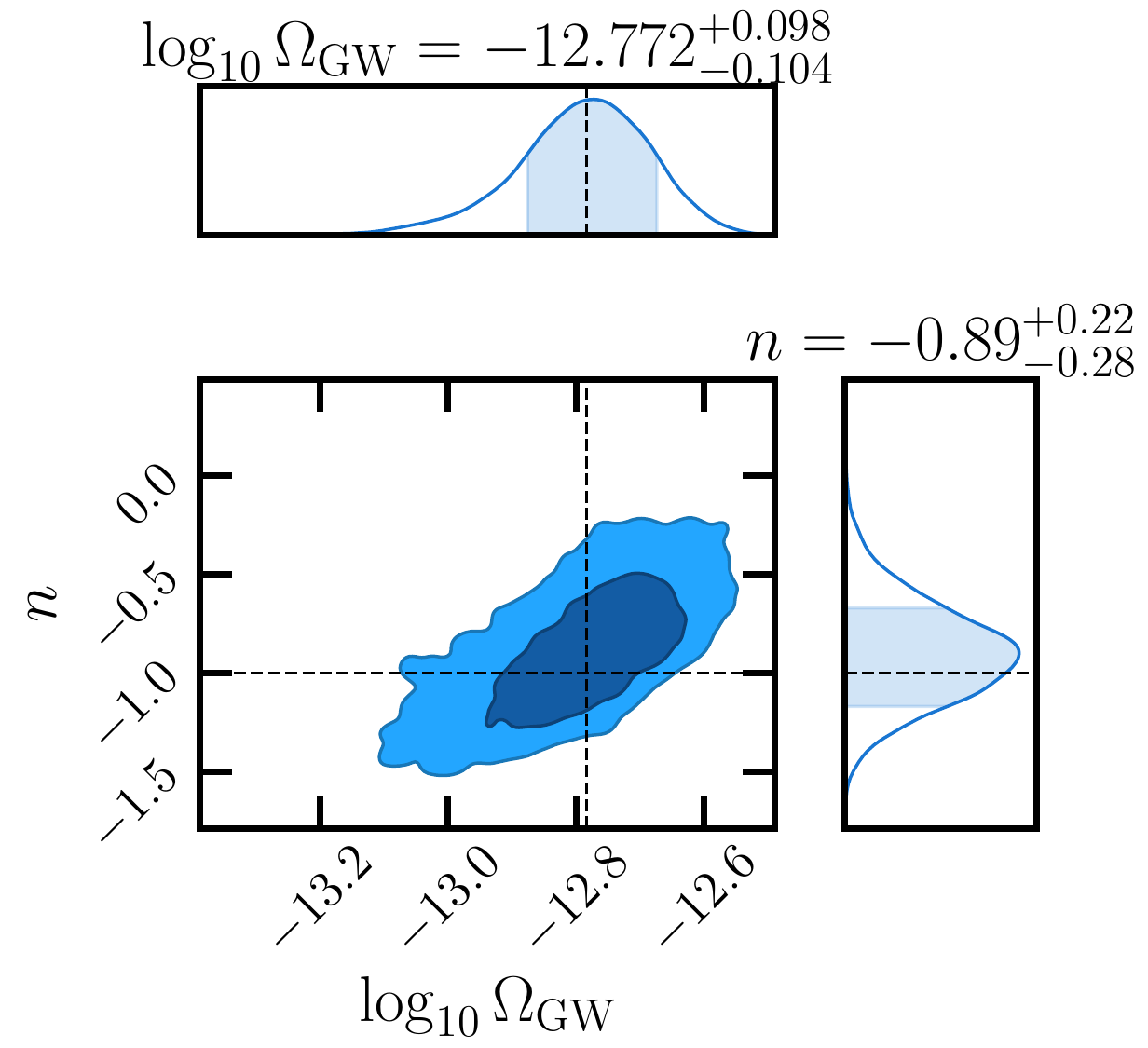}
	\caption{Posteriors of the \gls{sgwb} parameters (log-energy density and power-law index) for an injection of $\Omega_{0} = \num{1.63E-13}$ and $n = -1$.}
	\label{fig:posterior_sgwb_1e-13_n-1}
\end{figure}

\begin{figure}[ht]
	\centering
	\includegraphics[width=\columnwidth, trim={0 0.5cm 0 0.5cm}, clip]{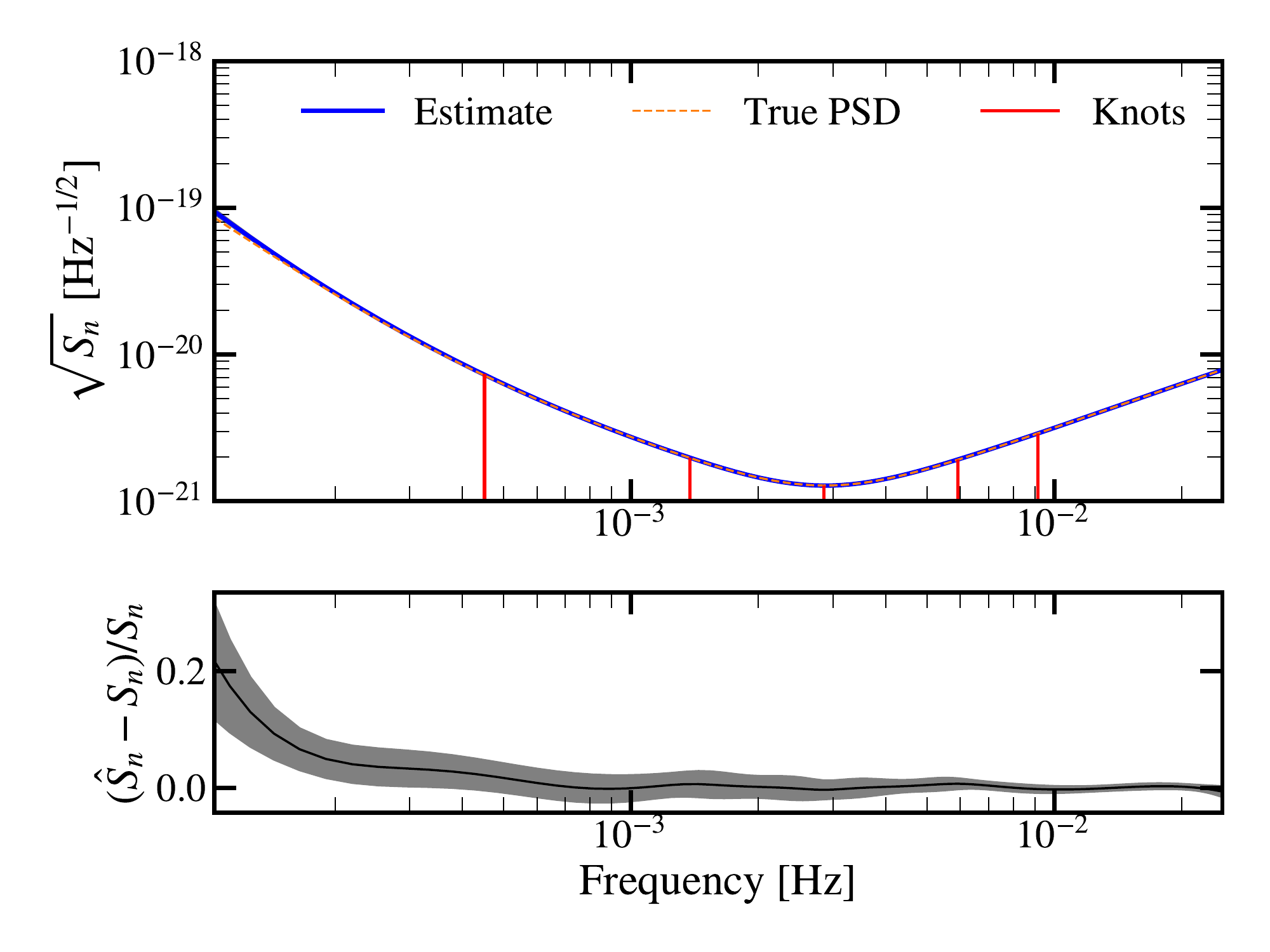}
	\caption{Upper panel: posterior of the single-link noise \gls{psd} (blue) compared to the true value (dashed orange). The vertical red lines locate spline control points. Bottom panel: average relative error obtained with the \gls{map} estimate, along with the 3-$\sigma$ credible interval.}
	\label{fig:noise_posterior_1e-13_n-1}
\end{figure}

As an example of parameter posterior, we pick the case $\Omega_{0} = \num{1.63E-13}$ and $n = -1$. It is particularly interesting because it lies in the detection limit and also features a \gls{sgwb} strain \gls{psd} slope of $-4$, which is similar to the low-frequency shape of the strain sensitivity curve (in power). We plot the signal parameters' joint posterior in~\cref{fig:posterior_sgwb_1e-13_n-1} and verify that the injected values lies within the credible interval. We also compute the corresponding \gls{tdi} signal and noise \glspl{psd} from posterior samples in \cref{fig:posterior_1e-13_n-1}. The \gls{map} of the \gls{gw} signal parameters yields the red solid curve, which is close to the true \gls{psd} shown by the dashed purple curve, even though the credible interval is relatively large. The noise \gls{psd} represented by the blue curve is better constrained as it dominates over the signal in the entire frequency band. This is confirmed by the spline reconstruction of the links' noise \gls{psd} in \cref{fig:noise_posterior_1e-13_n-1}, where the \gls{map} estimate (in blue) coincides with the true \gls{psd} (dashed orange) with a relative error smaller than \SI{10}{\percent} in most of the analyzed frequency band.

\subsection{Validity of the averaged Bayes factors}
\label{sec:verification}

In this section, we check that the averaged Bayes factor $\overline{\log \mathcal{B}_{10}}$ we compute with the method outlined in \cref{sec:detection} is consistent with what we obtain with single data realizations. We generate simulated datasets following the model described in \cref{sec:simu}; we include different realizations of both the noises and the \gls{sgwb} for a handful of cases.

\begin{figure}[ht]
	\centering
	\includegraphics[width=\columnwidth, trim={1.0 0.5cm 1.5 0.5cm}, clip]{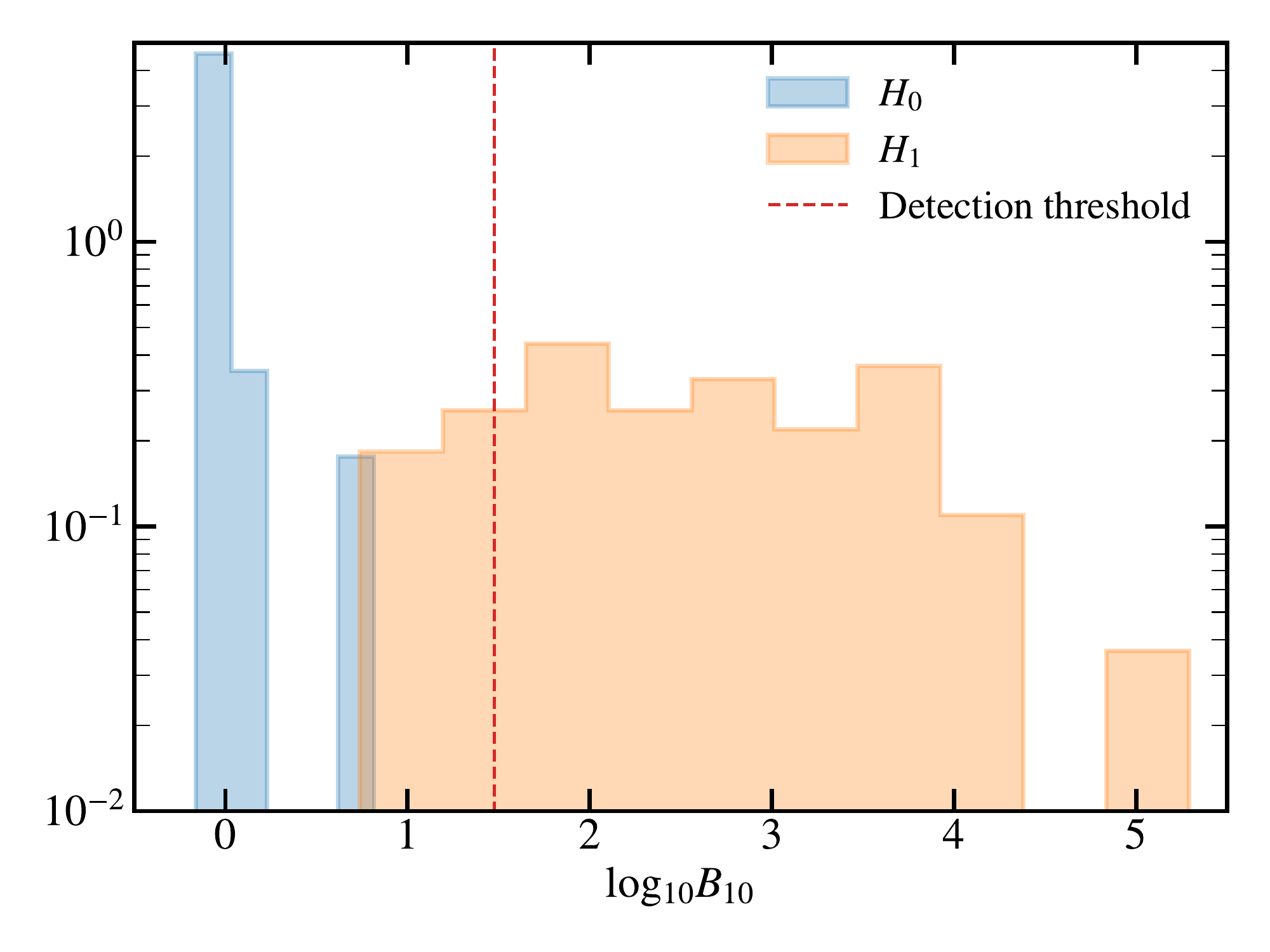} 
	\caption{Statistics of the decimal log-Bayes factor for couples of parameters $(\log \Omega_{0}, n)$ corresponding to the detection threshold $\mathcal{B}_{10} = 30$ (vertical red dashed line) derived from the contour plot in \cref{fig:bayes_factor_contours}. For each power-law index, Bayes factors are computed for 10 data realizations under $H_1$ (noise and signal, in blue). The histogram of log-Bayes factors computed for 20 data realizations under $H_0$ (noise only) is also shown in orange.}
	\label{fig:bayes_factor_statistics}
\end{figure}

As we are particularly interested in \gls{lisa}'s ability to detect a \gls{sgwb} as a function of its shape, we extract the pairs of parameters defining the contour line $\overline{\mathcal{B}_{10}}(\Omega_0, n) = 30$ corresponding to the detection threshold (dashed orange line in \cref{fig:bayes_factor_contours}). For each of these pairs corresponding to an integer power law index between $n=-2$ and $n=3$, we generate 10 data realizations under hypothesis $H_1$, from which we sample the posterior distributions and compute the evidences under both $H_0$ and $H_1$. We plot the histogram of the log-Bayes factors we obtain in \cref{fig:bayes_factor_statistics} (orange), along with the detection threshold line (dashed red). The distribution we obtain exhibits a significant variance, but the mean is located towards Bayes factor values larger than the threshold. Among the Bayes factors estimated from these simulations, \SI{80}{\percent} yield a value above the detection threshold.

In addition, we perform a similar analysis with 30 data realizations generated under hypothesis $H_0$ (containing only noise), and plot the histogram of the log-Bayes factors we obtain in blue on the same figure. They are concentrated around zero and distributed approximately like a chi-squared distribution. All the simulations produce values below the detection threshold, i.e., there are no false positive for these data realizations. The orange and blue distributions show that our derivation of detection limit is a conservative one as it minimizes the false-alarm rate at the expense of \SI{20}{\percent} of false negatives.

\section{Conclusion}
\label{sec:discussion}

We have presented a method to detect \glspl{sgwb} from \gls{lisa} measurements, which, for the first time, is model-agnostic with respect to the instrumental noise spectral shape. Instead, we use a flexible model for the single-link noise \glspl{psd} based on cubic splines. Such modelling could avoid biasing the instrument characterization and the subsequent impact on the signal detection. We test for the presence of an isotropic \gls{sgwb} through Bayesian model comparison, where we model both the signal and the noise transfer functions. We also adopt a template-based search to look for power-law signals. As a step towards more realistic instrumental setup compared to previous studies, we simulate interferometric data in the time domain, featuring a spacecraft constellation with unequal, time-varying armlengths. In this configuration, the assumptions underlying classic pseudo-orthogonal \gls{tdi} variables $A, E, T$ break down. Therefore, we directly analyze the three second-generation Michelson variables $X_2$, $Y_2$, $Z_2$ and account for their full frequency-dependent covariance matrix. We restrict the observation time to one year and the analyzed frequency bandwidth to the interval \SIrange{0.1}{50}{\milli\hertz} to mitigate computation time and artefacts related to blind frequency spots of \gls{lisa}'s sensitivity.

We run multiple injections of \glspl{sgwb} with a wide range of energy densities and power-law spectral indices to determine the region of the parameter space that would allow for a detection. We confirm \gls{lisa}'s ability to detect a scale-invariant \gls{sgwb} with an energy density above $\sim \num{2E-13}$, a threshold that was previously reported in the literature, in spite of the added flexibility on the noise modeling. This confirms \gls{lisa}'s ability to detect \glspl{sgwb} that not accessible to today's \gls{gw} detectors. In addition, we show that with a pivot frequency of $f_{0} \sim \SI{3}{\milli\hertz}$ and power-law indices ranging between $n=-2$ and $n=3$, we can distinguish \gls{gw} backgrounds from noise provided that their \gls{snr} is sufficiently large. We also probe larger absolute values of indices, keeping in mind that such extreme cases are unlikely to correspond to any signal as they would have been detected by current observatories.

This work motivates further investigations to improve the robustness of \glspl{sgwb} searches with space-based observatories against instrumental noise modeling. In this perspective, future works will account for distinct transfer functions for the different noise sources, and in particular for acceleration and readout noises. We also plan to allow for different noise levels across the various interferometers. Moreover, we performed our study based on a power-law model of isotropic stochastic signals, which does not reflect the full diversity of processes that can lead to stochastic backgrounds of \glspl{gw}. We plan to test other templates, but also to assess to what extent one can be agnostic with respect to both the signal and noise shapes while preserving the ability to tell them apart. As a final step, we aim to include the various astrophysical stochastic signals in our analysis, thus testing this pipeline to the greater LISA global fit scheme~\cite{Littenberg2023xpl}.

\acknowledgments
The authors thank the LISA Simulation Expert Group for all simulation-related activities. They would like to personally thank J. Veitch for their insightful feedbacks. J.-B.B. gratefully acknowledges support from UK Space Agency (grant ST/X002136/1).  N.K. acknowledges support from the Gr-PRODEX 2019 funding program (PEA 4000132310). Some of the results in this paper have been derived using the healpy and HEALPix package.

\appendix

\section{Derivation of the time-domain response function}
\label{app:derivation-response}

\begin{figure}
	\centering
	\begin{tikzpicture}[scale=4,tdplot_main_coords]
		
		\coordinate (O) at (0,0,0);
		\draw[thick,->] (0,0,0) -- (1,0,0) node[anchor=north east]{$\vb{x}$};
		\draw[thick,->] (0,0,0) -- (0,1,0) node[anchor=north west]{$\vb{y}$};
		\draw[thick,->] (0,0,0) -- (0,0,1) node[anchor=south]{$\vb{z}$};
		
		\tdplotsetcoord{P}{\rvec}{\thetavec}{\phivec};
		\draw[thick,color=red] (O) -- (P);
		\draw[dashed, color=red] (O) -- (Pxy);
		\draw[dashed,color=red] (P) -- (Pxy);
		
		\tdplotdrawarc{(O)}{0.2}{0}{\phivec}{anchor=north}{$\lambda_k$}
		\tdplotsetthetaplanecoords{\phivec}
		\tdplotdrawarc[tdplot_rotated_coords]{(0,0,0)}{0.3}{\thetavec}{90}{anchor=south west}{$\beta_k$};
		
		\tdplotsetcoord{Pth}{\rvec}{\thetavec - 12}{\phivec};
		\tdplotsetcoord{Pph}{\rvec}{\thetavec}{\phivec - 20};
		\tdplotsetcoord{Pr}{\rvec - 0.3}{\thetavec}{\phivec};
		\draw[-stealth, color=blue] (P)--(Pth) node[above] {$ \vu{v}_k = -\vu{e}_{\theta,k}$};
		\draw[-stealth, color=blue] (P)--(Pph) node[below left] {$\vu{u}_k = -\vu{e}_{\phi,k}$};
		\draw[-stealth, color=blue] (P)--(Pr) node[below right] {$\vu{k}_k = -\vu{e}_{r,k}$};
		
		\fill (P) circle[radius=0.5pt];
		\draw[dashed,tdplot_rotated_coords] (1.0,0,0) arc (0:90:1.0);
		\draw[dashed] (1.0,0,0) arc (0:90:1.0);
		\draw[dashed] (1,0,1.19) arc (0:60:1.0);
		
	\end{tikzpicture}
	\caption{Parametrization of the localization for point source $k$. The propagation vector is $\mathbf{\hat{k}_k}$, and the polarization vectors are $\mathbf{\hat{u}_k}$ and $\mathbf{\hat{v}_k}$. Adapted from the LDC Manual, available at \url{https://lisa-ldc.lal.in2p3.fr}. }
	\label{fig:ssb-frame}
\end{figure}
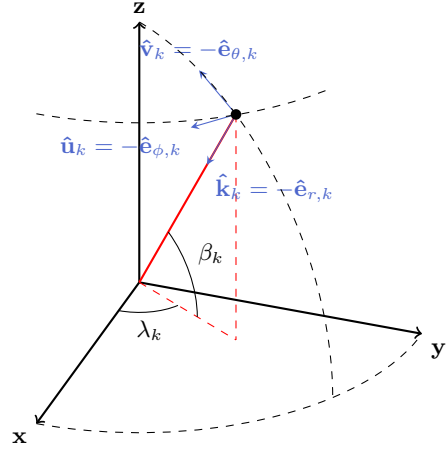

We express each stochastic point source's position using the Cartesian coordinate system $(\vb{x}, \vb{y}, \vb{z})$, defined such that $(\vb{x}, \vb{y})$ is the plane of the ecliptic. We introduce the associated spherical coordinates $(\theta, \phi)$, based on the orthonormal basis vectors $(\vu{e}_r, \vu{e}_\theta, \vu{e}_\phi)$, as illustrated in \cref{fig:ssb-frame}. The $k$-th source localization is parametrized by the \textit{ecliptic latitude} $\beta_k = \pi / 2 - \theta_k$ and the \textit{ecliptic longitude} $\lambda_k = \phi_k$. The basis vectors read
\begin{subequations}
	\begin{align}
		\vu{e}_{r,k} &= (\cos \beta_k \cos \lambda_k, \cos \beta_k \sin \lambda_k, \sin \beta_k), \\
		\vu{e}_{\theta,k} &= (\sin \beta_k \cos \lambda_k, \sin \beta_k \sin \lambda_k, -\cos \beta_k), \\
		\vu{e}_{\phi,k} &= (-\sin \lambda_k, \cos \lambda_k, 0).
	\end{align}
\end{subequations}

The propagation vector is $\vu{k}_k = -\vu{e}_{r,k}$. We define the \textit{polarization vectors} as $\vu{u}_k = -\vu{e}_{\phi,k}$ and $\vu{v}_k = -\vu{e}_{\theta,k}$. This produces, for source $k$, a direct orthonormal basis $(\vu{u}_k, \vu{v}_k, \vu{k}_k)$.

The time series of frequency shifts $y_{12,k}(t)$, experienced by light traveling along link $12$, is computed by projecting the strain of point source $k$ on the link unit vector $\vu{n}_{12}$ (computed from the spacecraft positions),
\begin{equation}
	\begin{split}
		H_{12,k}(t) ={}& h_+(t, \vu{n}_k) \xi_+(\vu{u}_k, \vu{v}_k, \vu{n}_{12})
		\\
		&+ h_\times(t, \vu{n}_k) \xi_\times(\vu{u}_k, \vu{v}_k, \vu{n}_{12}),
	\end{split}
	\label{eq:projected-strain-ssb}
\end{equation}
where we assume that the link unit vector $\vu{n}_{12}$ is constant during the light travel time. The \textit{antenna pattern functions} are given by
\begin{subequations}
	\begin{align}
		\xi_+(\vu{u}_k, \vu{v}_k, \vu{n}_{12}) &= \qty(\vu{u}_k \vdot \vu{n}_{12})^2 - \qty(\vu{v}_k \vdot \vu{n}_{12})^2,
		\\
		\xi_\times(\vu{u}_k, \vu{v}_k, \vu{n}_{12}) &= 2 \qty(\vu{u}_k \vdot \vu{n}_{12}) \qty(\vu{v}_k \vdot \vu{n}_{12}).
	\end{align}
\end{subequations}

Light emitted by spacecraft 2 at $t_2$ reaches spacecraft 1 at $t_1$. Accounting for the effect of source $k$ only, these two times $t_1$ and $t_2$ are related by $H_{12,k}(\vb{x}, t)$,
\begin{equation}
	t_1 \approx t_2 + \frac{L_{12}}{c} - \frac{1}{2c} \int_0^{L_{12}}{H_{12,k}(\vb{x}(\lambda), t(\lambda)) \dd{\lambda}},
	\label{eq:reception-time-with-H}
\end{equation}
We approximate the wave propagation time to first order as $t(\lambda) \approx t_2 + \lambda/c$. Also, $\vb{x}(\lambda) = \vb{x}_2(t_2) + \lambda \vu{n}_{12}(t_2)$, where $\vb{x}_2(t_2)$ represents the position of the emitter spacecraft at emission time. Using these two expressions, we can further refine $H_{12,k}$ as
\begin{align}
	\begin{split}
		&H_{12,k} \qty(\vb{x}(\lambda), t(\lambda)) 
		= H_{12,k} \qty(t(\lambda) - \frac{\vu{k}_k \vdot \vb{x}(\lambda)}{c})
		\\
		&= H_{12} \qty(t_2 - \frac{\vu{k}_k \vdot \vb{x}_2(t_2)}{c} + \frac{1 - \vu{k}_k \vdot \vu{n}_{12}(t_2)}{c} \lambda),
	\end{split}
	\label{eq:link-deformation-function-of-x}
\end{align}
Combining \cref{eq:link-deformation-function-of-x,eq:reception-time-with-H} and differentiating the resulting expression with respect to $t_2$ yields the relative frequency shift, $y_{12}$, experienced by light as it travels along link 12,
%\begin{widetext}
	\begin{align}
		y_{12,k}(t_2) & \approx \frac{1}{2 \qty(1 - \vu{k}_k \vdot \vu{n}_{12}(t_2))}  \nonumber \\
		& \left[ H_{12,k} \qty(t_2 - \frac{\vu{k}_k \vdot \vb{x}_2(t_2)}{c}) \right.  \nonumber   \\
		& \left. - H_{12,k} \qty(t_2 - \frac{\vu{k}_k \vdot \vb{x}_1(t_1)}{c} + \frac{L_{12}}{c}) \right].
	\end{align}
	Here, we have introduced the receiver spacecraft position at reception time $\vb{x}_1(t_1) = \vb{x}_2(t_2) + L_{12} \vu{n}_{12}(t_2)$. These spacecraft positions are expressed in the coordinate frame introduced represented \cref{fig:ssb-frame}, and computed with \texttt{LISA Orbits}~\cite{lisaorbits}.
	
	Using $t_1 \approx t_2 + L_{12} / c$ and the fact that the spacecraft moves slowly compared to the propagation timescale, we obtain $\vb{x}_2(t_2) \approx \vb{x}_2(t_1)$ and $\vu{n}_{12}(t_1) \approx \vu{n}_{12}(t_2)$,
	\begin{align}
		y_{12,k}(t_1) & \approx \frac{1}{2 \qty(1 - \vu{k}_k \vdot \vu{n}_{12}(t_1))}  \nonumber \\
		& \left[ H_{12,k} \qty(t_1 - \frac{L_{12}(t_1)}{c} - \frac{\vu{k}_k \vdot \vb{x}_2(t_1)}{c}) \right. 
		\nonumber \\ 
		& \left. - H_{12,k} \qty(t_1 - \frac{\vu{k}_k \vdot \vb{x}_1(t_1)}{c}) \right],
	\end{align}
%\end{widetext}
where the equation for $y_{12,k}$ is now solely a function of reception time $t_1$. Finally, combining \cref{eq:y-sum,eq:projected-strain-ssb,eq:instrument-response-to-point-source} gives $y_{12}$ as a function of $t_1$ in terms of the point sources' strains.

\section{Derivation of the stochastic gravitational-wave background response in the frequency domain}
\label{sec:sgwb_response_freq}

In this section, we derive the frequency-domain covariance of two links due to an isotropic and stationary \gls{sgwb} given by \cref{eq:link-gw-covariance}.

The measured response to a particular polarization $p=+,\times$ includes the contribution from all sky locations, so that
\begin{equation}
	y_{lm, p}(t) = \int_{\vu{k}}{y_{lm, p}(t, \vu{k}) \dd[2]{\vu{k}}}.
	\label{eq:y_integral}
\end{equation}
We can obtain the expression for $y_{lm, p}(t, \vu{k})$ by combining \cref{eq:instrument-response-to-point-source} and \cref{eq:projected-strain-ssb} to get
\begin{align}
	\begin{split}
		y_{lm, p}(t, \vu{k}) \approx{}& \frac{1}{2 \qty(1 - \vu{k} \vdot \vu{n}_{lm}(t))}
		\left[ \right. \\
		& \left.  h_{p}\qty(t - \frac{L_{lm}(t)}{c} - \frac{\vu{k} \vdot \vb{x}_m(t)}{c}, \vu{n}_k) \right. \\
		&- \left. h_{p}\qty(t - \frac{\vu{k} \vdot \vb{x}_l(t)}{c}, \vu{n}_k) \right] \xi_{p}(\vu{u}_k, \vu{v}_k, \vu{n}_{lm}).
	\end{split}
	\label{eq:instrument-response-analysis}
\end{align}
Then, we decompose the time-domain \gls{gw} perturbation $h_{p}(\tau, \vu{k})$ on the Fourier basis as 
\begin{equation}
	h_{p}(\tau, \vu{k}) = \int_{-\infty}^{+\infty}{\tilde{h}_{p}(f, \vu{k}) e^{2 \pi i f \tau} \dd{f}}.
\end{equation}
Injecting this decomposition into \cref{eq:instrument-response-analysis} yields
\begin{align}
	\begin{split}
		y_{lm, p}(t, \vu{k}) \approx \int_{-\infty}^{+\infty}{\tilde{h}_{p}(f', \vu{k}) e^{2\pi i f' t} G_{lm, p}(f', t, \vu{k}) \dd{f'}},
	\end{split}
	\label{eq:instrument-response-kernel}
\end{align}
where we defined the kernel
\begin{align}
	\begin{split}
		G_{lm, p}(f', t, \vu{k}) ={}& \frac{\xi_{p}(\vu{u}_k, \vu{v}_k, \vu{n}_{lm}) }{2 \qty(1 - \vu{k} \vdot \vu{n}_{lm}(t))}
		\Big[ \\
		& e^{ - \frac{2\pi i f'}{c} \qty(L_{lm}(t) + \vu{k} \vdot \vb{x}_m(t))}  - e^{- \frac{2\pi i f'}{c} \vu{k} \vdot \vb{x}_l(t)} \Big].
	\end{split}
\end{align}
Now we compute the Fourier transform of \cref{eq:instrument-response-kernel} evaluated at frequency $f$, which yields
\begin{equation}
	\tilde{y}_{lm, p}(f, \vu{k}) = \int_{-\infty}^{+\infty} \tilde{h}_{p}(f', \vu{k}) \tilde{G}_{lm, p}(f', f - f', \vu{k}) \dd f',
	\label{eq:instrument-response-kernel-fourier}
\end{equation}
which is the convolution of the gravitational strain with the Fourier transform of the kernel
\begin{equation}
	\label{eq:kernel-fourier}
	\tilde{G}_{lm, p}(f', f, \vu{k}) \equiv \int_{-\infty}^{+\infty} G_{lm, p}(f', t, \vu{k}) e^{-2 \pi i f t} \dd{t}.
\end{equation}
For isotropic, stationary, zero-mean backgrounds with \gls{psd} $S_h$, the strain covariance can be written as
\begin{equation}
	\label{eq:strain-covariance}
	\mathrm{E}\qty[\tilde{h}_{p}(f, \vu{k}) \tilde{h}^{\ast}_{p'}(f', \vu{k}')] = \frac{1}{8\pi} S_{h}(f) \delta(f - f') \delta(\vu{k} - \vu{k}') \delta_{p p^{\prime}}.
\end{equation}

Let us label the covariance of two links $lm$ and $l'm'$ as
\begin{equation}
	\label{eq:link-covariance-definition}
	C_{lm, l'm', p}(f) \equiv \mathrm{E}\qty[\tilde{y}_{lm, p}(f), \tilde{y}^{\ast}_{l'm', p}(f)].
\end{equation}
Plugging \cref{eq:y_integral} and \cref{eq:instrument-response-kernel-fourier} into \cref{eq:link-covariance-definition}, owing to isotropy and stationarity we obtain
\begin{align}
	\begin{split}
		C_{lm, l'm', p}(f) & = 
		\iint_{\vu{k}} S_h(f')\tilde{G}_{lm, p}(f', f - f', \vu{k})  \\
		& 
		\tilde{G}^{\ast}_{l'm', p}(f', f - f', \vu{k}) \dd{f'} \dd[2]{\vu{k}}.
	\end{split}
	\label{eq:link-covariance}
\end{align}
The above expression can be simplified by noting that \gls{lisa}'s response to a infinitely large number of incoherent sources (a background) only very weakly depends on time (up to about \SI{1}{\percent}), although the response to a \gls{gw} with wave vector $\vu{k}$ has time variations. In other words, sky averaging washes out the time dependence, so that one can approximate the averaged response at $t$ by its value at any given time $t_0$. As a result, we can write \cref{eq:link-covariance} as the product of the strain \gls{psd} and a response function that directly depends on the time-domain kernel,
\begin{equation}
	C_{lm, l'm', p}(f) = S_h(f) R_{lm, l'm', p}(f, t_0),
	\label{eq:link-covariance-fixed-time}
\end{equation}
where we defined 
\begin{equation}
	R_{lm, l'm', p}(f, t_0) \equiv \int G_{lm, p}(f, t_0, \vu{k}) G^{\ast}_{l'm', p}(f, t_0, \vu{k}) \dd[2]{\vu{k}}.
	\label{eq:response_matrix_elements}
\end{equation}
This equation allows us to compute the elements of the link response matrix involved in \cref{eq:link-gw-covariance}, after summing over the two polarizations.

%% References
%\bibliographystyle{apsrev4-1}
%\bibliographystyle{unsrt85}
% \bibliographystyle{unsrt}

\printglossary
\bibliography{references}

\end{document}